\documentclass[twocolumn,showpacs,preprintnumbers,amsmath,amssymb]{revtex4}
\usepackage{graphicx,epsfig,amsfonts}
\usepackage{psfrag}

\begin{document}

\title{Complex population dynamics \\
as a competition between multiple time-scale phenomena}

\author{Ioana Bena}
\email{ioana.bena@physics.unige.ch}
\affiliation{University of Geneva, Theoretical Physics Department, 
Quai E. Ansermet no. 24, 1211 Geneva 4,
Switzerland}
\author{Michel Droz}
\email{michel.droz@physics.unige.ch}
\affiliation{University of Geneva, Theoretical Physics Department, Quai E.
Ansermet no. 24, 1211 Geneva 4,
Switzerland}
\author{Janusz Szwabi\'{n}ski}
\email{janusz.szwabinski@physics.unige.ch}
\affiliation{University of Geneva, Theoretical Physics Department, Quai E.
Ansermet no. 24, 1211 Geneva 4,
Switzerland \\
and Institute of Theoretical Physics, University of Wroc{\l}aw,
Pl. M. Borna 9, 50-204 Wroc{\l}aw, Poland}
\author{Andrzej P\c{e}kalski}
\email{apekal@ift.uni.wroc.pl}
\affiliation{Institute of Theoretical Physics, University of Wroc{\l}aw,
Pl. M. Borna 9, 50-204 Wroc{\l}aw, Poland}

\date{\today}

\begin{abstract}

The role of the selection pressure and mutation amplitude on the behavior of 
a single-species population evolving on a two-dimensional lattice, 
in a periodically changing environment, is studied
both analytically and numerically.
The mean-field level of description allows to highlight the delicate interplay
between the different time-scale processes in the resulting complex dynamics of
the system. We clarify the influence of the amplitude and period of the
environmental changes on the critical value of the selection pressure
corresponding to a phase-transition ``extinct-alive" of the population.
However, the intrinsic stochasticity and  the dynamically-built in correlations 
among the individuals,
as well as  the role of the mutation-induced variety in population's 
evolution are not appropriately
accounted for. A more refined level of description, which is an
individual-based one, has to be considered.
The inherent fluctuations do not destroy the phase transition
``extinct-alive", and the mutation amplitude is strongly influencing
the  value of the critical selection pressure. The phase diagram in the 
plane of the population's parameters -- selection and mutation is discussed 
as a function of the environmental variation characteristics. The differences 
between a smooth variation of the environment and an abrupt, 
catastrophic change are also addressesd. 
\end{abstract}


\maketitle


\section{Introduction}
\label{introduction}

The study of population dynamics is a recurrent
subject of interdisciplinary research ever since
the pioneering works of Lotka~\cite{lotka} and 
Volterra~\cite{volterra}. Most of the situations
involve interacting multi-agent systems, that
evolve in a spatially-extended environment, and 
exhibit complex cooperative behavior, 
see e.g.~\cite{complex1,complex2}.
Such systems are generally far from equilibrium 
and highly nonlinear.  Moreover, it has been 
recognized that simple mean-field (MF) like approaches 
are not always able to grasp all the richness 
of the behavior of such systems~\cite{granada05}. 
The stochastic aspect of the dynamics, 
as well as the use of discrete 
spatial and/or temporal variables to characterize 
the population evolution may play 
an important role~\cite{bradshaw,durrett}. Also,
the inherent discrete nature of a population, i.e., 
the fact that it consists of a finite, integer number
of individuals, was found to be sometimes an influencial
factor of the dynamics, 
see~\cite{henson1,domokos2}. 

Therefore, the methods developed in nonequilibrium 
statistical physics are particularly well-suited for 
the study of such systems.
In particular, the so-called 
{\em individual-based model} (IBM) approach 
is able to account for all the above-cited 
ingredients of the dynamics.

A major part of the existing literature is investigating
various evolution mechanisms in a {\em constant} environment,
like, e.g., Darwinian selection leading 
to the  elimination of the ill-adapted individuals~\cite{dar}, 
genetic heritage and accumulation of diseases~\cite{he}, 
mutations~\cite{stauffer}, development of hunting/avoiding strategies 
of predators/preys~\cite{my}, topology 
of the network connecting the individuals~\cite{net}, etc. 
As a result of the 
nonlinear stochastic dynamics associated with 
the above mechanisms, 
nontrivial collective properties are emerging. 
In particular, depending on the values of the 
control parameters 
characterizing a system, different steady states 
are possible~\cite{us} and a phase diagram can be drawn.
 
However, a fundamental question concerning the dynamics 
of populations or ecosystems is
their robustness with respect to external perturbations, 
like, for example, a  change in their environment. 
Such an environmental change can be abrupt 
(due, e.g., to some catastrophic events) 
or rather smooth and periodic 
(due, e.g., to slow climatic modifications).  
Although the climatic 
changes recorded in paleoclimatic data~\cite{bradley} 
could hardly be called perfect oscillations, one 
notices a certain degree of periodicity.  Periodic 
changes of the habitat are also assumed in most of the
studies done both by biologists and physicists, see
e.g.~\cite{pease,lande,sznajd},
and can be regarded as a rather 
satisfactory approximation 
to the real data, as far as extracting 
essential {\em qualitative} features is
concerned.

One expects on general grounds that new complex 
behavior of the populations will emerge due to 
the competition between 
the characteristic intrinsic time scales and nonlinearities
of the system and the time scale of the 
environmental perturbation. This is an aspect that,
to our knowledge, has not been investigated systematically
in the existing literature. Indeed, most of the previous studies
on related points (see, e.g.,~\cite{granada05,holt,burger})
are only reporting on the observed complex dynamics, without
further analysis of its rooting.
In this paper we shall thus address the role 
of selection pressure and mutation amplitude 
on the relevant properties of a single-species population 
which lives in a periodically changing habitat. 
Their interplay with the magnitude and period of 
the environmental changes will be a fundamental point 
of the analysis.

The paper is organized as follows: in the following section we 
present the details of the model; its implementation, 
both in a MF-like and an IBM-type of
approach, are described in Sec.~\ref{implement}. 
Sec.~IV presents the analytical and numerical results obtained
in the MF approximation.
Section~V resumes the main results of the 
IBM simulations, and compares them to the
results of the MF approach. 
Finally, conclusions and perspectives are relegated to Sec.~VI. 
 

\section{The model}
\label{model}

We consider a population that consists of 
{\em hermaphrodite} individuals  living 
in a {\em spatially-extended} habitat, 
which is represented here as a two-dimensional 
square lattice 
of size $L \times L$. 
We assume non-permeable boundary conditions, 
meaning that the individuals cannot cross
the borders of the lattice. Moreover, the lattice 
has a {\em finite carrying capacity}, which comes
from an exclusion assumption, according to which
there is at most one individual in each lattice node.
 
The dynamics of the population is the result of several 
elements, namely natural selection, individual motion, 
mating and reproduction, whose modeling is described below.\\ 

\noindent{\em A. Natural selection. Individual 
trait, optimum, fitness, selection pressure, extinction probability.}\\  
Each individual is characterized by its {\em trait}, 
or phenotype, which may correspond to various specific 
properties, such as e.g. the metabolic rate,
the body size, or may encode a rather complex
ensemble of properties. 
The trait of the ``$i$-th" individual
is represented here, for simplicity,
through a real number  $z_i \in$ [0,1]. 
This number  is 
 fixed  at the 
individual's birth, and remains constant.
This simplified way to characterize 
an individual has been often used both by biologists~\cite{burger, holt} 
and by physicists~\cite{mdap2}. Its comparison with more elaborated
descriptions, that take into account further details of the 
genetic structure of the individuals, was addressed, for example, in~\cite{krzys}.

The population lives in an environment
whose influence on the individuals is  
encoded in the value of another number 
$\varphi \,\in\,  [0,1]$, the so-called {\it optimum}.
The optimum is that value of the trait which guarantees for its 
owner maximum chance of survival.
In principle, the optimum could 
be spatial and/or 
time-dependent. In this work the optimum has
the same value in all the sites of the lattice, 
however its value can vary in time.

The degree of agreement of the individual 
trait $z_i$ with the optimum  determines 
the individual {\em extinction probability
per unit time}  (or {\em extinction rate}),
\begin{equation}
p_i\,=p_0\left[1-\,\exp\left(- \frac{{\cal S}}{f_i}\right)\right]\,,
\label{surv}
\end{equation}
where ${\cal S}$ is a parameter which models 
the {\em selection pressure} (SP)
of the environment  and constitutes
a main  control parameter of the system. 
The {\it fitness} $f_i$ of the individual $i$ is defined as
\begin{equation}
f_i = 1 - |z_i - \varphi|\,.
\label{fitness}
\end{equation}
A perfectly-adapted individual has a fitness equal to $1$, and thus a 
minimum possible extinction rate.
A fitness smaller than $1$ corresponds to
a worse adaptation;
$f_i$ goes to zero for the completely unadapted individuals, that have the
maximum possible extinction rate, equal to $p_0$. 
A variation in time of the fitness
of one individual is resulting only from the variation of the optimum.

The constant $p_0$ in the definition~(\ref{surv}) 
of the extinction rate is related to the choice 
of the  unit time and it depends on whether we are considering
the continuous-time MF-type model, or the discrete-time
IBM-type of approach, see below.

The choice we made of the extinction rate~(\ref{surv}),  
the implicit definition of the SP parameter ${\cal S}$ 
and that of the fitness~(\ref{fitness})  
are frequently encountered
in the biological literature, see e.g.~\cite{burger}. Other choices, 
and thus other ways of measuring the ``selection pressure" and the
``individual fitness", are possible. However, most of them can be 
mapped one onto the other and/or account for equivalent {\em qualitative} 
aspects of  the interaction between the individuals and their environment. \\

\noindent{\em B. Individual motion.}\\
If an individual survives, it can
move to its surroundings.
The simplest possibility, that we shall adopt hereafter,
is a random-walk, diffusive-like motion. 
For example, in the discrete-time IBM-type of approach,
in one time step the individual jumps on the lattice, 
from its initial location
to a randomly chosen nearest-neighbor one  (i.e., a site 
within the von Neumann neighborhood of the initial node), 
provided that the chosen site is empty,
and that it lies within the boundaries of the system. If none of the 
four first-neighbor nodes is
empty, then the individual cannot move, 
and thus cannot mate (see below).\\

\noindent{\em C. Mating and reproduction. Heredity and mutation.}\\ 
Suppose an individual $i$ reaches a destination node.  
Then, if there are other individuals (``neighbors") 
in the nearest-neighborhood of this
destination site, the individual ``i" choses at random 
one of these neighbors (call it ``$j$") 
for mating~\footnote{Although hermaphrodite,
the individuals need mating for reproduction.}. 
The pair of individuals $i$ and $j$ may afterwards give birth to 
offsprings, whose number cannot exceed a prescribed
value  $N_{{\rm off}}$.
At their birth, the progenies are placed, at random,
on the empty nodes of the joint nearest-neighborhoods of 
the two parents (that counts $6$ sites); 
therefore, the maximum number of 
offsprings $N_{{\rm off}}\leqslant 6$.
If there is no room in this neighborhood for putting 
an offspring, then this one is not born.

The trait of a progeny $k$ coming from parents $i$ and $j$ 
is determined by the parents' traits (heredity), 
but it can also present some ``variations" due to different random factors,
 such as recombination,  mutations, 
 etc. We shall consider
\begin{equation}
z_k = \frac{1}{2} (z_i + z_j) + m_k \,,
\label{zoff}
\end{equation}
where $m_k$  represents these variations.
It brings diversification into the phenotypic 
pool of the population and we call it conventionally {\em mutation}. 
For simplicity, we shall admit that $m_k$ is a random number,
uniformly distributed in the interval $[-{\cal M},\,{\cal M}]$, 
where $0<{\cal M}<1$ is called hereafter the {\em mutation amplitude} (MA)
and is  a control parameter of the system~\footnote{In the biological
literature parameters analogous to ${\cal M}$ are often referred to as {\em
mutation rate}. However, from the point of view of the physical aspect it
designates, the term {\em mutation amplitude} looks more appropriate to us.}.
Moreover, if Eq.~(\ref{zoff}) leads to 
$z_k >1$ or $z_k<0$, then one renormalizes it
by resetting it to $z_k=1, 0$, respectively.
This means simply that the trait of the individuals cannot 
overcome some fixed limits.
This choice~(\ref{zoff}) for the trait of
an offspring is often made in the biological literature~\cite{burger}.

The population dynamics is thus driven by two main ``forces"
that are acting, to some extent, 
in opposite directions: selection and mutation, 
characterized, respectively, 
through the values of the control
parameters ${\cal S}$ and ${\cal M}$.
 Selection, combined with heredity, tries to 
bring the average trait close to the instantaneous optimum, 
while mutation introduces diversity in the 
individual traits, and thus is broadening the 
distribution of the population's traits.


\section{Implementations of the model}
\label{implement}

The study of this system 
can be made on different levels of modelling.
The simplest one, which has the advantage to 
allow, up to a certain extent, 
for analytic investigations, 
is a MF-like level. The drawback of 
this simple approach is to neglect fluctuations, 
which can  play a crucial role. 
We are therefore proposing an investigation
based on IBM-type simulations, for which the
fluctuations are naturally built-in. A comparison of the
results of these two approaches is then made.

\subsection{Mean-field like approach}
\label{mf1}

The main quantity of interest in this approach,
that neglects the possible correlations 
between lattice sites, is
the average population number density $c(t)$.
One writes down a continuous-time evolution equation 
for  $c(t)$ by taking 
into account the different dynamical processes described 
above, 
\begin{equation}
\frac{dc(t)}{dt}=c\left[R_1(c)R_2(c)\,-p\,\right]\,.
\label{mfeq}
\end{equation}
This represents the balance between births,
described by the nonlinear term $[c\,R_1(c)R_2(c)]$, and deaths,
corresponding to the $(-c\,p)$ term.
The meaning of these terms is as following:\\
(a) $p$ is the mean extinction probability per unit time, 
and, as resulting from
Eqs.~(\ref{surv}), (\ref{fitness}),  it reads 
\begin{equation}
p=p_0\left[1-\exp\left(-\frac{{\cal S}}{1-|\varphi-\langle z\rangle|}\right)\right]\,,
\label{mfprob}
\end{equation}
where $\langle z \rangle$  is the mean value of the trait.
The time scale will be chosen such that the constant
$p_0=1$.

In agreement with Eq.~(\ref{zoff}),
the MF approach does not allow for a self-consistent 
evaluation of  the mean trait $\langle z \rangle$; it
remains thus an arbitrary constant  $\in\,[0,\,1]$. 
The role of the mutation is practically
eliminated in the MF description, and the
only remaining control parameter is the selection ${\cal S}$.\\
(b) $R_1(c)$ 
expresses the probability that an individual 
jumps, per unit time, to a randomly-chosen 
empty nearest-neighbor site, 
\begin{equation}
R_1(c)=r_0\left(1-c^4\right)\,,
\end{equation}
where $(1-c^4)$ is simply the probability that 
 at least one of these four neighbor sites is empty.
The constant $r_0$ depends on the choice of the unit 
time. If  the unit time is fixed by chosing
$p_0=1$, then $r_0$ is also fixed, and it measures the 
ratio between the mean survival time of an individual and
its characteristic diffusion time on the lattice.\\
(c) The term $R_2(c)$ describes the probability that the considered
individual  encounters at least one nearest-neighbor at the
destination site and it produces a certain number of
offsprings according to the model rules:
\begin{equation}
R_2(c)=N_{{\rm off}}\,c\,(1-c)^{N_{{\rm off}}}\,.
\end{equation}
One is supposing implicitely that the time scale related to mating and 
reproduction is much shorter than both the time scales of survival and
diffusion of an individual.

It can be easily shown that $r_0 \gg 1$ implies a rapid increase of $c$
till the stationary saturation value $c=1$, while $r_0 \ll 1$
leads to an extinction of the population. In the foregoing we shall thus
focus on the case when diffusion and extinction have comparable 
characteristic time scales, and we shall fix throughout $r_0=1$.

One is then led to the following MF equation:
\begin{eqnarray}
\frac{dc(t)}{dt}&=&N_{{\rm off}}\,c^2\,(1-c^4)(1-c)^{N_{{\rm off}}}\nonumber\\
&-&
c\,\left[1-\exp\left(-\frac{{\cal S}}{1-|\varphi-\langle z\rangle|}\right)\right]\,.
\end{eqnarray}
This one  can be mapped 
onto the equation of motion of an overdamped
particle in a potential $U_{{\rm sel}}(c)$ whose 
actual profile depends on the selection parameter ${\cal S}$,
\begin{equation}
\frac{dc(t)}{dt}=-\,\frac{dU_{{\rm sel}}(c)}{dc}\,.
\label{potential}
\end{equation}
This point of view will prove to be very useful in the 
discussion of the results presented in Sec.~\ref{mf2}.

\subsection{IBM-type simulations}
\label{ibm1}

The IBM simulation algorithm considers the individuals
distributed on the lattice nodes, the initial condition being 
represented by  their  positions and the prescribed values 
of the individual traits. 
Different initial conditions were considered, see Sec.~\ref{ibm2} below. 
The individuals are evolving, at discrete Monte-Carlo time steps (MCS,
defined hereafter), according to the stages~{\em A--C} of the dynamics as
described in the Introduction, namely:\\
 {\em A.} At a given time $t$ 
an individual $i$ is picked at random,
and its extinction probability  $p_i$, 
corresponding to one MCS, is determined according to 
Eqs.~(\ref{surv}),  (\ref{fitness}).  The constant $p_0$
in Eq.~(\ref{surv}) is chosen equal to $1$. 
Then a random number $r$ is extracted from
an uniform distribution in the range $[0,\,1]$; if $r<p_i$, the individual
dies, otherwise it survives.\\
{\em B.} If it survives, the individual $i$ jumps at random to one of the empty
nearest-neighbor nodes on the lattice.\\
{\em C.} Then it possibly mates and produces offsprings.

If at the time  $t$ there are $N(t)$ individuals in the system, 
then the above steps {\em A--C} are repeated $N(t)$ times; 
this constitutes one MCS, the unit-time of the simulations.
Finally, the time is advanced by one step, $t \rightarrow t+1$, 
and the above
algorithm is repeated.

In the next two Sections we shall present and 
then compare the results obtained
at each of these two levels of modeling.

\section{Mean-field type analysis}
\label{mf2}

\subsection{Constant environment}
\label{mf2const}

Consider first the case of a habitat with constant 
optimum $\varphi=0.5$, and a population described by a mean 
trait $\langle z \rangle =0.5$~\footnote{One should notice that 
changing one or both of the values of
$\varphi$ and $\langle z \rangle$ is simply equivalent to a rescaling of the 
selection parameter ${\cal S}$ 
in Eq.~(\ref{mfprob}).}. We shall suppose throughout
that $N_{{\rm off}}=6$.
\begin{figure}
\psfrag{sel}{{\large$\;\;{\cal S}$}}
\begin{center}\quad \vspace{0.5cm}\\
\includegraphics[width=0.9\columnwidth]{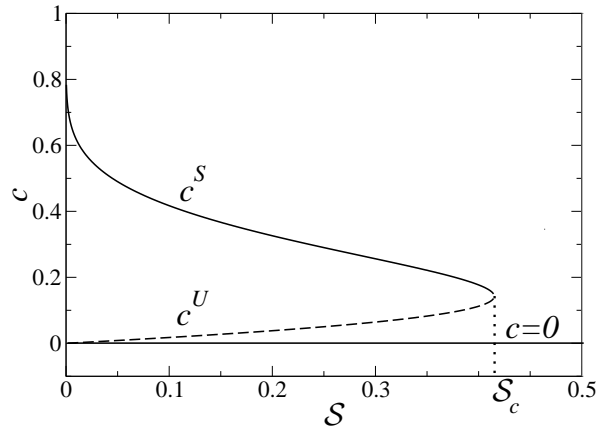}
\caption{\label{figure1}
MF stationary solutions for the concentration $c$ 
as  functions of the selection parameter ${\cal S}$.
Continuous lines represent the stable solutions $c=c^S$ and $c=0$, 
and the dashed
line represents the unstable solution $c=c^U$. At 
${\cal S}={\cal S}_c \approx 0.417$ the stable $c^S$ and unstable
$c^U$ solutions disappear, and for ${\cal S} > {\cal S}_c$ 
the only remaining stationary
solution is the absorbing state $c=0$.
The other parameters are $\varphi=\langle z\rangle=0.5$, and $N_{{\rm off}}=6$.}
\end{center}
\end{figure}

One can use simple arguments of the theory of deterministic
dynamical systems to describe the evolution of the
mean population concentration $c(t)$.
The stationary concentrations
as  functions of the 
unique control parameter ${\cal S}$ are represented in Fig.~\ref{figure1}.
At a critical value of the selection ${\cal S}={\cal S}_c\approx 0.417$
a discontinuous phase transition  
takes place in the system. 
For ${\cal S}<{\cal S}_c$, the system presents two stable states, namely
$c^S\neq 0$  and the absorbing state $c=0$, and  their respective basins of
attraction are separated by the unstable solution $c^U$.
At ${\cal S}={\cal S}_c$ the stable $c^S$ and unstable $c^U$ 
critical points of the overdamped dynamics~(\ref{mfeq})
disappear. For large
${\cal S}>{\cal S}_c$ the population gets extinct
in the long-time limit, no matter what 
was its initial concentration $c(0)$.

The standard linear stability analysis~\cite{nicolis} 
allows also to determine 
the characteristic 
times for the linear relaxation to the stable point $c^S$, 
(denoted $T_{{\rm rel}}^S$) and, 
 for the escape from the 
immediate vicinity of the unstable point $c^U$,
denoted by $T_{{\rm esc}}^U$.
One can also get the analytic expression for the linear
relaxation time to the absorbing state $c=0$, namely
$T_{{\rm rel}}^0=\left[1-\exp(-{\cal S})\right]^{-1}$.
These characteristic times offer an indication on the
order of magnitude of the nonlinear relaxation times
that correspond to the fully nonlinear dynamics~(\ref{potential}).
Moreover, as shown numerically, 
for a wide domain of intermediate values of the 
selecion pressure $0<{\cal S}<{\cal S}_c$, they are practically insensitive
to changes in ${\cal S}$.
This key aspect 
is implicitely taken into account 
in discussing the cases (i) and (ii) in Sec.~IVB.

Another point of interest for the non-trivial stable 
stationary solution $c^S$ is its
dependence on the maximum number of offsprings $N_{{\rm off}}$. 
One can retain two important aspects, namely:\\
(a) At low concentrations (where the exclusion hypothesis is practically irrelevant), 
the survival in a highly-demanding environment (large $\mathcal S$)
is more efficiently ensured when more descendants are born at each mating.\\
(b) In a medium- and low-demanding
environment, $c^S$ is increasing with decreasing $N_{{\rm off}}$ 
(for a given value of ${\cal S}$).
Indeed, a population that gives birth to fewer descendants at each mating 
has to compensate the death rate (that is fixed by ${\cal S}$) 
through more mating, and thus
through a larger concentration. However, this concentration cannot exceed a
certain value, since, on one hand, the individuals have to move in order to
mate (and the corresponding jump probability $R_1(c)$ is decreasing as $1-c^4$), 
and, on the other
hand, there has to be ``enough room" for the born offsprings to be 
put-down (indeed, the corresponding probability $R_2(c)$ is rapidly-decaying as 
$\approx\,(1-c)^{N_{{\rm off}}}$ for large values of $c$).  
The  stationary value  $c^S$ results thus from the balance between these
contradictory tendencies.

\subsection{Time-periodic habitat}
\label{mf2osc}

Suppose that the environment is changing periodically in time,
and  consider, as the simplest possibility:
\begin{equation}
\varphi(t)=0.5+A \sin\left(2 \pi \;\frac{t-t_{{\rm i}}}{T_p}\right) 
\,\Theta(t-t_{{\rm i}})\,.
\label{varopt}
\end{equation}
Here $A$ denotes the amplitude of the environmental perturbation 
($0 < A\leqslant 0.5$), $T_p$ is its period, and $t_{{\rm i}}$ 
is the moment of
onset of the perturbation; $\Theta$ is the Heaviside step function.
The mean trait of the population is
still supposed constant, $\langle z \rangle =0.5$, and the 
selection has a certain given value ${\cal S}={\cal S}_{{\rm i}}$. 

This situation is formally equivalent
to that of a population of constant mean trait in a constant
environment, $\langle z \rangle = \varphi = 0.5$, but for which the
 ${\cal S}$ oscillates, with a period $T_p/2$, as:
\begin{equation}
{\cal S}(t)=\frac{{\cal S}_{{\rm i}}}{1-A\left|\sin\left(2 \pi \;
\displaystyle\frac{t-t_{{\rm i}}}{T_p}\right)\right|} 
\,\Theta(t-t_{{\rm i}})\,.
\label{varS}
\end{equation}
This point of view allows us to explain easily the behavior of the concentration 
depending on the value of ${\cal S}_{{\rm i}}$ and
on the parameters $A$ and $T_p$ of the perturbation.
\begin{figure}
\begin{center}\quad \vspace{0.5cm}\\
\includegraphics[width=0.9\columnwidth]{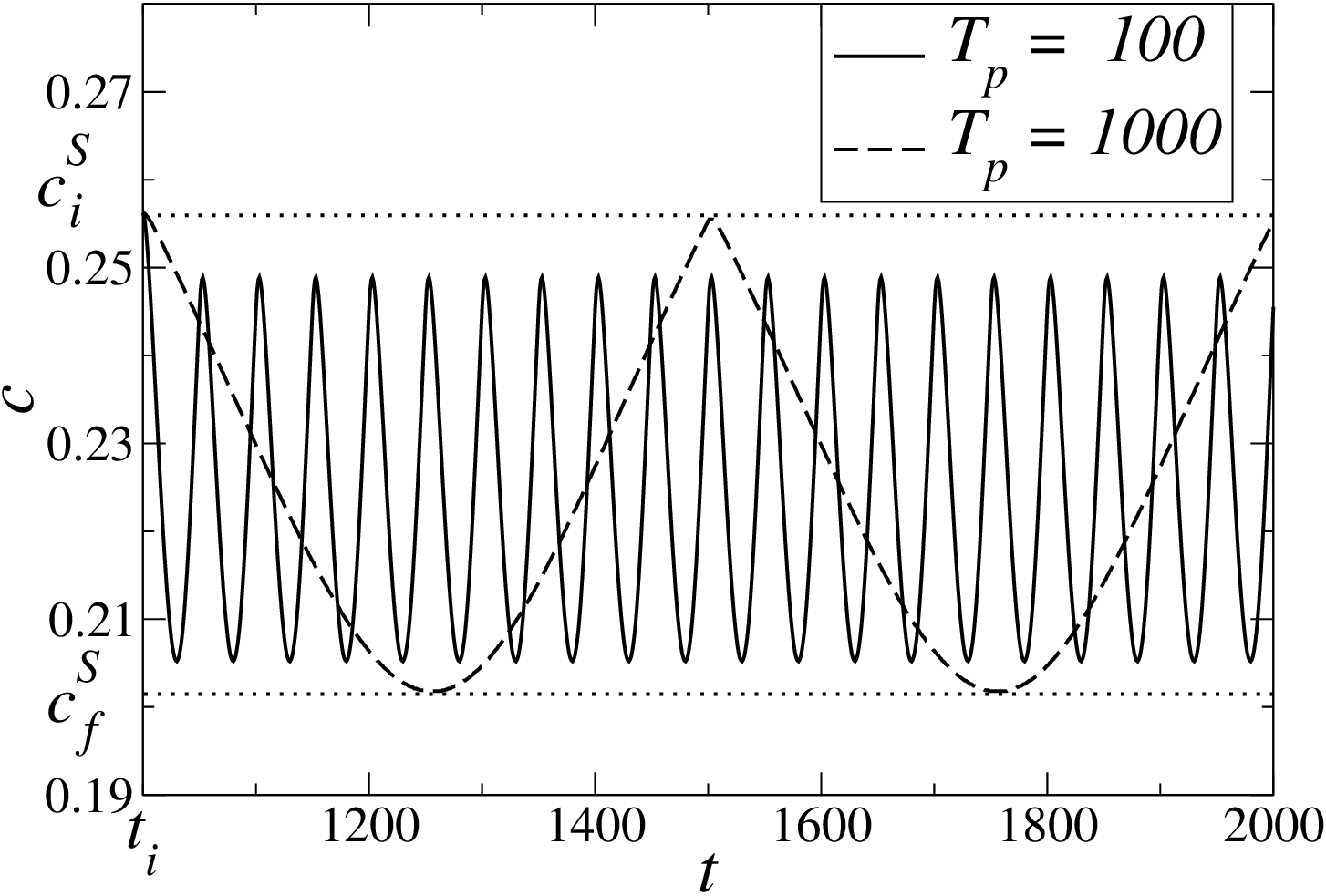}\quad\vspace{1.cm}\\
\includegraphics[width=0.9\columnwidth]{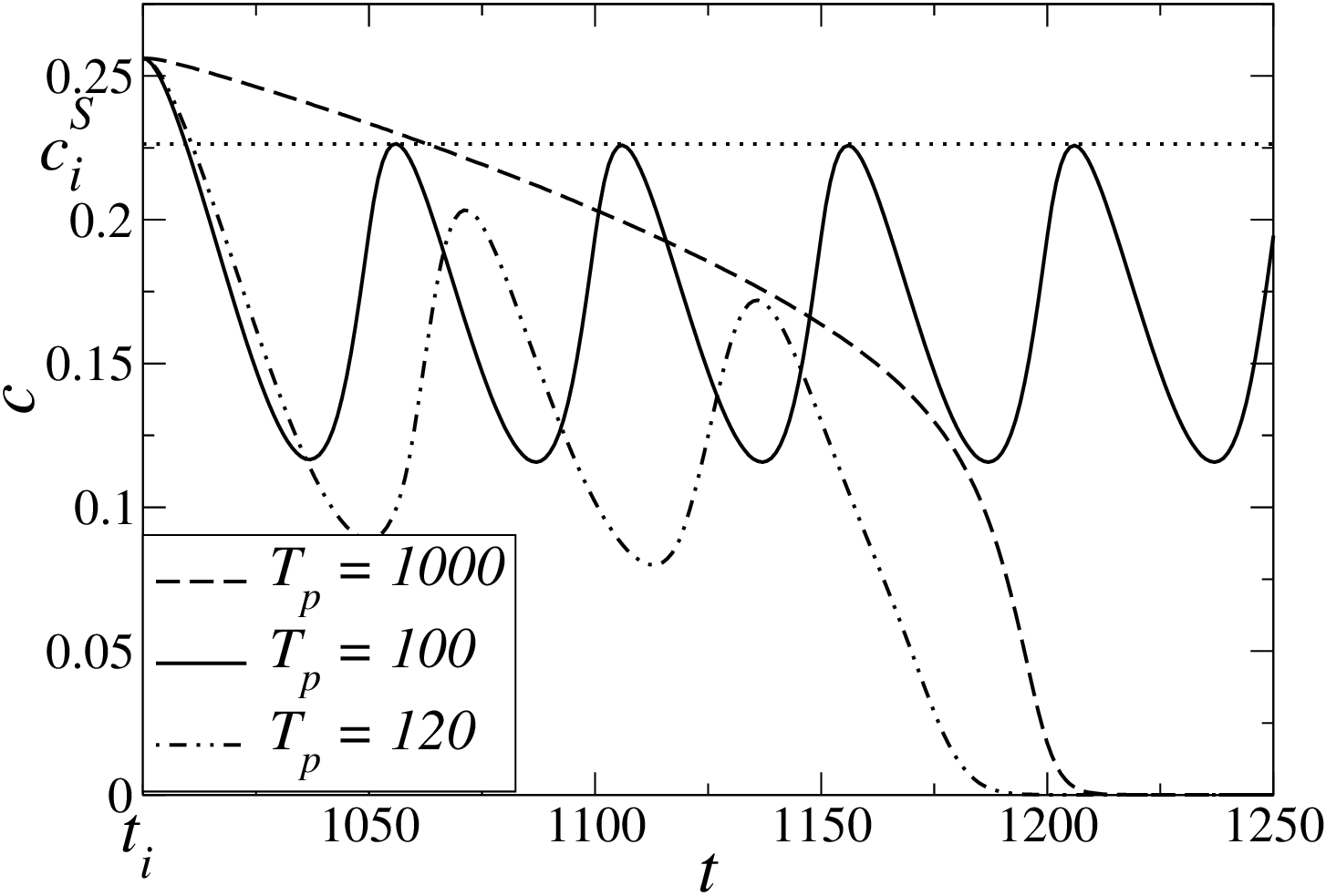}
\caption{\label{figure5} 
MF: Possible temporal evolutions of the concentration $c(t)$ in a
periodically-changing habitat
for different values of $A$ and of $T_p$.
The other parameters are $c(0)=0.7$,
${\cal S}_{{\rm i}}=0.3$ and $t_{{\rm i}}=1000$.
The upper panel refers to the
{\em case (i)} described in the  text and is obtained for $A=0.2$.
The lower panel corresponds to the {\em case (ii)} in the text, 
and is obtained for $A=0.35$.}
\end{center}
\end{figure}

For ${\cal S}_{{\rm i}}\geqslant {\cal S}_c$, the population
dies after a time of the order of $T_{{\rm rel}}^0$,
so this case is not particularly interesting. 

Let us suppose that ${\cal S}_{{\rm i}}<{\cal S}_c$, and that $c(0)$ is in the
basin of attraction of the stable solution $c^S_{{\rm i}}$
that corresponds to ${\cal S}_{{\rm i}}$.
After a time of the order
of $T_{{\rm rel}}^S$, the concentration relaxes to $c^S_{{\rm i}}$.
We suppose
that the perturbation of the selection parameter is applied after reaching this state, i.e.,  
$t_{{\rm i}} \gg T_{{\rm rel}}^S$. After that,  ${\cal S}(t)$ starts
to grow, and at $t_i+T_p/4$ it is reaching the final maximum value 
${\cal S}_{{\rm f}}={\cal S}_{{\rm i}}/(1-A)$; 
then it decreases during $T_p/4$ till reaching the
initial value ${\cal S}_{{\rm i}}$, etc.
Two situations may appear:\\

\noindent{\em Case (i): ${\cal S}_{{\rm f}} <{\cal S}_c$}.\\ 
The amplitude $A$ of the perturbation is such that
${\cal S}(t)$ never overcomes ${\cal S}_c$.
Then the nontrivial attractor $c^S$ (with $c^S\leqslant c^S_{{\rm i}}$) never 
ceases to exist, and $c$
always remains in its basin of attraction. The population never dies, 
and its concentration is oscillating in time, with
a period $T_p/2$ and an amplitude that is generally determined by both $A$ and $T_p$,
see the upper panel of Fig.~\ref{figure5}.\\

\begin{figure}
\psfrag{X}{\large${\cal S}$}
\psfrag{X =0.3}{\tiny${\cal S}_i=0.3$}
\psfrag{X =0.34}{\tiny${\cal S}=0.34$}
\psfrag{X =0.375}{{\tiny{${\cal S}_f=0.375$}}}
\begin{center}\quad \vspace{0.5cm}\\
\includegraphics[width=0.9\columnwidth]{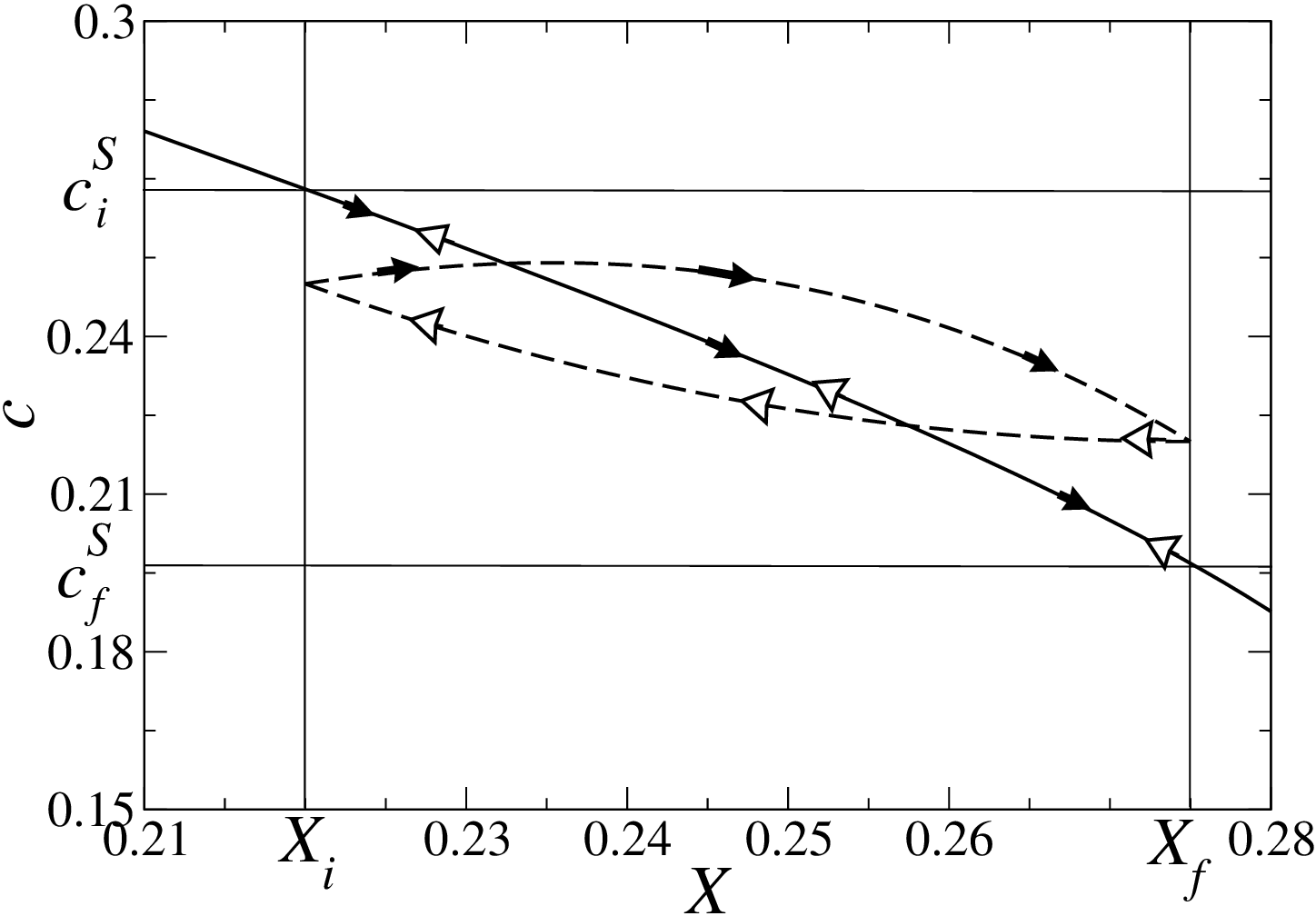}\quad\vspace{0.75cm}\\
\includegraphics[width=0.9\columnwidth]{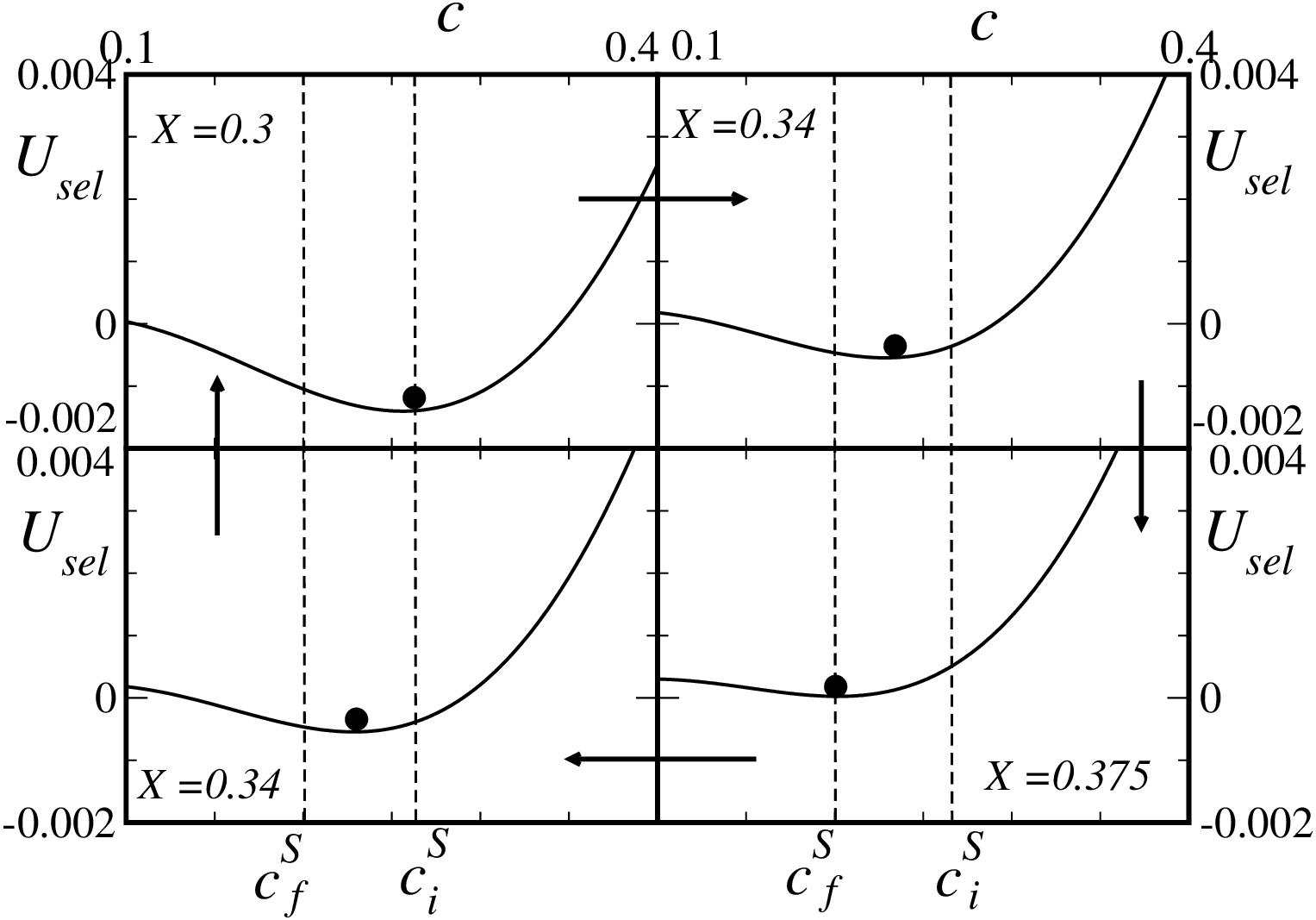}\quad\vspace{0.75cm}\\
\includegraphics[width=0.9\columnwidth]{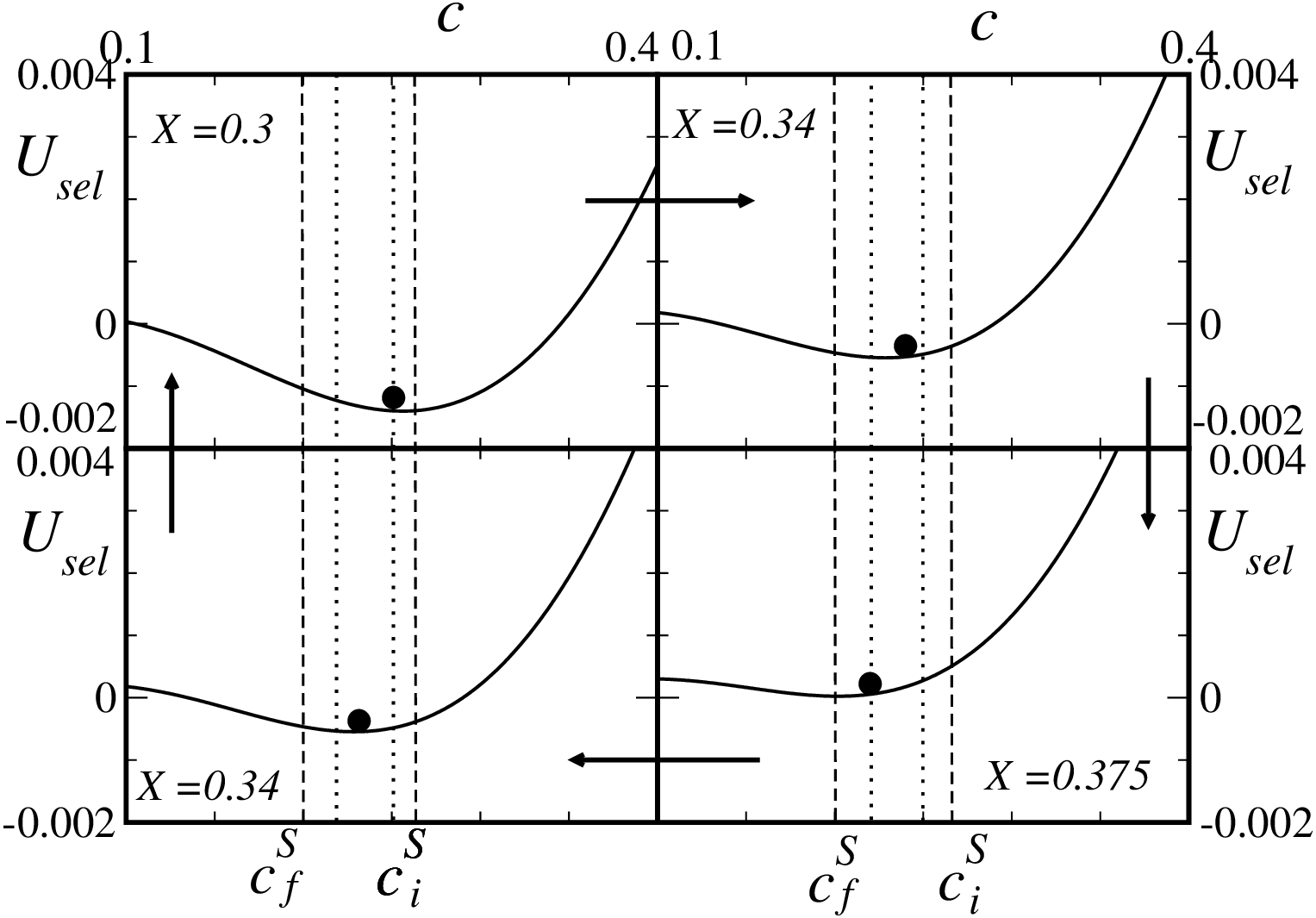}
\caption{\label{figure4i} MF: The upper panel 
describes schematically the 
trajectory of the system in the plane concentration-selection
for the {\em case (i)} as discussed in the text, for ${\cal S}_{{\rm i}}=0.3$ and 
$A=0.2$  (with $\varphi=\langle z\rangle =0.5$).
The continuous line corresponds to the {\em case (i.a)}
(slow perturbation), while the dashed line 
represents {\em case (i.b)} (rapid perturbation). 
Black arrows refer to the increase of ${\cal S}$, and
the white ones to the decrease of ${\cal S}$.
The middle panel illustrates the behavior of the representative point $c(t)$
(the big black dot) in terms of the oscillating potential $U_{{\rm sel}}(c)$
for {\em case (i.a)}, 
and the lower panel corresponds to {\em case (i.b)}. 
The thick arrows sketch the sense of evolution of $U_{{\rm sel}}(c)$; 
the values of ${\cal S}$
corresponding to the represented profiles of $U_{{\rm sel}}(c)$ are also indicated.}
\end{center}
\end{figure}

\noindent{\em Case (i.a)}. If the oscillations are very slow, $T_p \gg T_{{\rm rel}}^S$
for all the values of ${\cal S}_{{\rm i}}\leqslant {\cal S}\leqslant {\cal S}_{{\rm f}}$, then 
at each moment the concentration relaxes to the instantaneous value
of $c^S$, see the upper panel of Fig.~\ref{figure4i}. 
Thus, the concentration oscillates  between 
the two extremum possible values
$c^S_{{\rm i}}\geqslant c(t) \geqslant 
c^S_{{\rm f}}$ (where $c^S_{{\rm i,f}}$ 
are the equilibrium values corresponding to ${\cal S}_{{\rm i,f}}$).

This situation is also illustrated in the second panel of Fig.~\ref{figure4i}
in terms of the potential $U_{{\rm sel}}(c)$ that is driving the evolution of $c(t)$, see 
Eq.~(\ref{potential}). When ${\cal S}(t)$ is oscillating between 
${\cal S}_{{\rm i}}$ and ${\cal S}_{{\rm f}}$, the profile of the potential $U_{{\rm sel}}(c)$ 
is also changing periodically.
For a very slow evolution of $U_{{\rm sel}}(c)$, the representative point of the concentration
$c(t)$  is practically always ``stuck" to the minimum of the instantaneous potential.\\ 

\noindent{\em Case (i.b)} If the oscillations of the perturbation are rapid, 
i.e.,  $T_p$ is comparable to $T_{{\rm rel}}^S$,
then the concentration does not have time to relax 
to the instantaneous value of the attractor $c^S$, and the amplitude
of its oscillation is therefore smaller than in case (a), see the upper panel of Fig.~\ref{figure5}.
This amplitude is strongly dependent
on $T_p$, and for very small values of 
$T_p$ the concentration  
reaches practically a constant value. Thus,  for a 
very rapidly-changing
environment, one approaches a behavior that is reminiscent
of the one appearing in the limit of a constant
environment $A=0$.

The third panel of Fig.~\ref{figure4i} represents this situation in terms of the
competition between, on one hand, the relaxation of the representative point 
$c(t)$ to the instantaneous minimum 
of $U_{{\rm sel}}(c)$, and on the other hand,
the rapid variation of the  position of this minimum:
$c(t)$ lies always ``behind" the minimum.
\\

\noindent{\em Case (ii): ${\cal S}_{{\rm f}} >{\cal S}_c$ }.\\ 
Suppose that the amplitude $A$ of the perturbation allows the 
selection to overcome the critical value, i.e.,
${\cal S}_{{\rm f}}\geqslant {\cal S}(t) > {\cal S}_c$ during a certain 
interval of time $\tau$.
Then the attractor $c^S$ ceases to exist during 
this interval, and the concentration tends to relax
to the absorbing state $c=0$, on a time scale of the order
of $T_{{\rm rel}}^0$. \\

\begin{figure}
\psfrag{X}{\large${\cal S}$}
\psfrag{X =0.41}{\tiny${\cal S}=0.41$}
\psfrag{X =0.3}{\tiny${\cal S}_i=0.3$}
\psfrag{X =0.462}{{\tiny{${\cal S}_f=0.462$}}}
\begin{center}\quad \vspace{0.5cm}\\
\includegraphics[width=0.9\columnwidth]{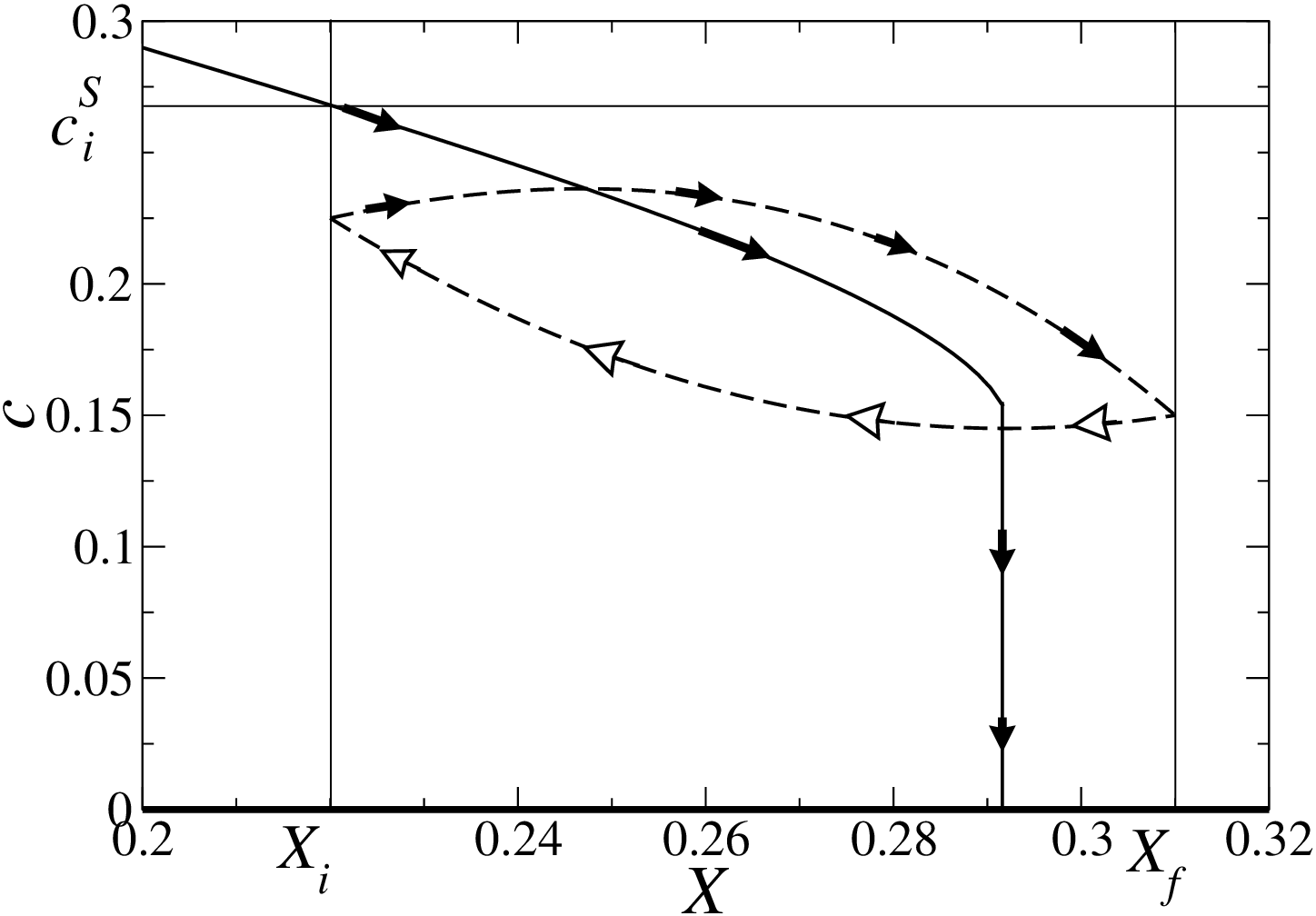}\quad\vspace{0.75cm}\\
\includegraphics[width=0.9\columnwidth]{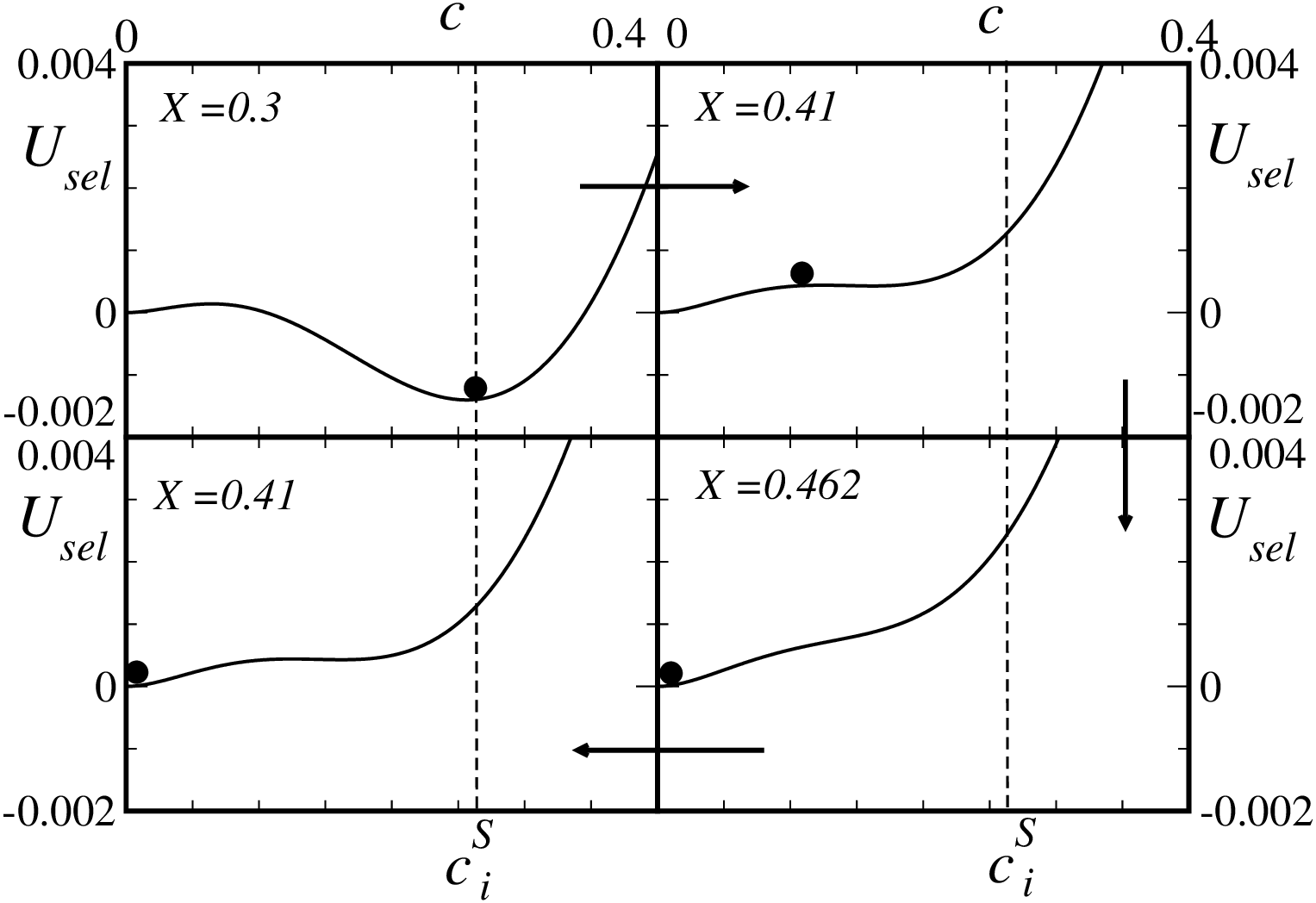}\quad\vspace{0.75cm}\\
\includegraphics[width=0.9\columnwidth]{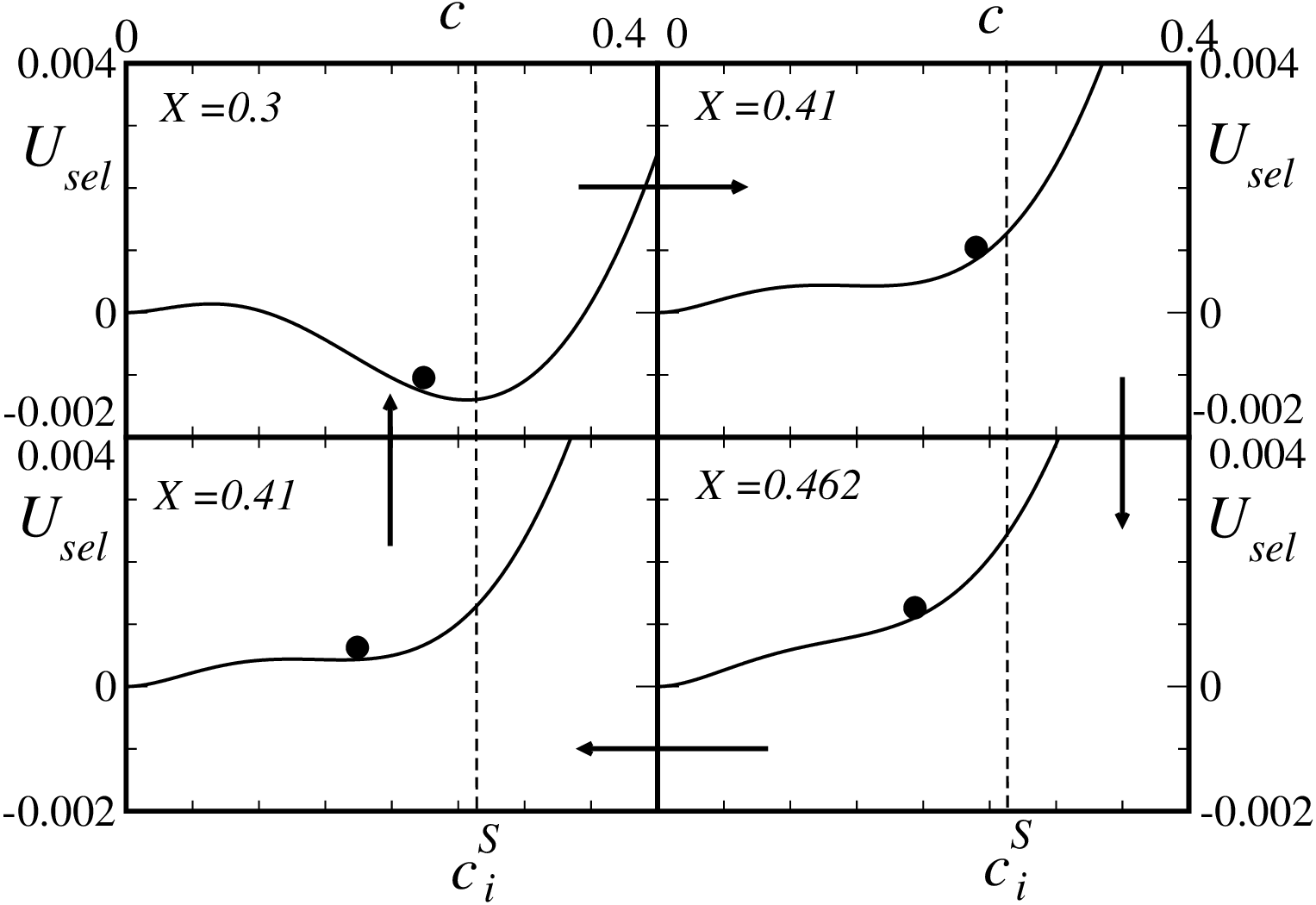}
\caption{\label{figure4ii}  MF: Upper panel: schematic representation of the 
trajectory of the system in the plane concentration-selection
for {\em case (ii)} of the main text,  for ${\cal S}_{{\rm i}}=0.3$ and 
$A=0.3$ (with $\varphi=\langle z\rangle =0.5$).
The continuous line corresponds to {\em case (ii.a)} (slow perturbation), and the dashed line 
represents {\em case (ii.b)} (rapid perturbation). 
Black/white arrows correspond to the increase/decrease of ${\cal S}$.
Middle and low panels:
the behavior of the representative point $c(t)$
(big black dot) when the profile of the potential $U_{{\rm sel}}(c)$ is evolving 
in the sense sketched by the thick black arrows.
Middle panel: {\em case (ii.a)}. Lower panel: {\em case (ii.b)}.
The values of ${\cal S}$ corresponding to the represented profiles 
of $U_{{\rm sel}}(c)$ are also indicated.}
\end{center}
\end{figure}

\noindent{\em Case (ii.a)} If $\tau \gg T_{{\rm rel}}^0$ (the case of a slow perturbation), 
then the population will die during the interval $\tau$. 
The corresponding evolution of the concentration 
is represented in the lower panel of Fig.~\ref{figure5},
see also the upper and middle panels of Fig.~\ref{figure4ii}
for an illustration.\\

\noindent{\em Case (ii.b)}. If, however, $\tau \lesssim T_{{\rm rel}}^0$ (rapid perturbation), 
then  the population may survive till the end of the interval $\tau$.
At this moment, the attractor $c^S$ starts to exist again;
if not already out of its basin of attraction, 
the concentration tends to relax to it, till ${\cal S}$ reaches again 
the initial value ${\cal S}_{{\rm i}}$ and the cycle starts all over again.
The behavior of the population in this case may be either
oscillating (after a transient regime), or may present damped 
oscillations to the absorbing state $c=0$ (the latter regime, however, is
appearing for extremely narrow intervals of the parameters). 
This is illustrated in the lower panel of Fig.~\ref{figure5}
and also schematically in the upper and lower panels 
of Fig.~\ref{figure4ii}.

\begin{figure}
\psfrag{sel}{{\;\;\large${\cal S}$}}
\psfrag{lll}{$\varphi=0.5$}
\psfrag{mmm}{$\varphi=0.8$}
\begin{center}\quad \vspace{0.5cm}\\
\includegraphics[width=0.9\columnwidth]{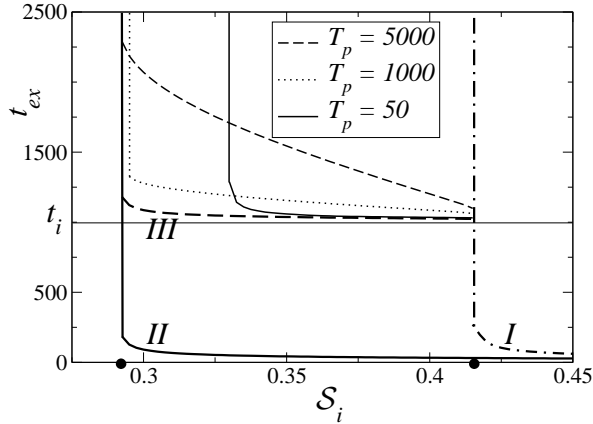}
\caption{\label{figure6} 
MF: Extinction time $t_{{\rm ex}}$ as a function of 
the initial selection parameter ${\cal S}_{{\rm i}}$ 
for different periods $T_p$ of the perturbation.
The amplitude of the perturbation is fixed to $A=0.3$. 
The critical values of ${\cal S}_{{\rm i}}$ are comprised between 
the extremal points 
${\cal S}_c(1-A)\approx 0.292$ and ${\cal S}_c \approx 0.417$ 
(represented by the big dots on the abcise). 
The continuous thick curve labelled
$I$ gives the ET as a function of ${\cal S}_i$ for
the case of a constant environment with $\varphi=0.5$, 
while the dashed-dotted  curve 
labelled $II$ is obtained for a constant environment
with $\varphi=0.8$. Finally, the thick dashed curve labelled $III$
corresponds to an abrupt jump in the optimum, from $\varphi=0.5$ 
for $t<t_i$ to
$\varphi=0.5+A=0.8$ at $t>t_i$.
The other parameters are
$\langle z\rangle =0.5$, $t_{{\rm i}}=1000$, and $c(0)=0.7$.}
\end{center}
\end{figure}

Another important characteristic 
of the system is the {\em extinction time} (ET), 
which represents the time it takes for the population  
to get extinct, i.e., to reach the absorbing $c=0$ state.
The phase transition ``extinct-alive" is reflected in the behavior of the
ET, which is finite for SPs  below 
the critical value and becomes infinite at 
the critical value of ${\cal S}_i$, as illustrated in Fig.~\ref{figure6}.

For a constant environment,
the ET is finite as long as 
${\cal S}_{{\rm i}}>{\cal S}_c\approx 0.417$, 
but it tends to infinity for ${\cal S}_{{\rm i}} <{\cal S}_c$ 
(provided, of course, that the initial concentration $c(0)$ is not 
in the basin of attraction of the absorbing state).

For the case of a time-periodic environment, the ET may also be 
infinite, or finite. 
In the first case the concentration $c(t)$ has an
oscillatory behavior, while in the latter it tends 
to the absorbing $c=0$ state.
The critical value of ${\cal S}_{{\rm i}}$ that separates these 
two situations depends in
general on both $A$ and $T_p$, and it is always 
comprised between ${\cal S}_c(1-A)$ and ${\cal S}_c$.
For very slow perturbations,
the critical value of ${\cal S}_i$ approaches
${\cal S}_c (1-A)$, which  is the one that 
corresponds to a constant
environment with $\varphi= 0.5+A$. 
Note, however, that increasing $T_p$
to the limit $T_p\rightarrow \infty$ (for a fixed $A$) does not 
imply that the general behavior of the
system ``converges" to the one corresponding to 
a constant environment with $\varphi=0.5+A$. This point is illustrated
in Fig.~\ref{figure6}: the profiles of ET corresponding to increasing $T_p$ 
values do not approach the profile for a constant environment with $\varphi=0.5+A=0.8$.

For rapid fluctuations the critical value of ${\cal S}_i$
depends strongly on $A$ and $T_p$. However, in the $T_p \rightarrow 0$ 
limit (infinitely-fast fluctuations),
it approaches  the value ${\cal S}_c$. More generally, the very fast
oscillations of the optimum (at arbitrary but fixed $A$) are ``smeared out",
simply because the relaxational dynamics of the system cannot follow these 
optimum fluctuations.
Thus the behavior of the system  converges to the one corresponding 
to the constant environment $\varphi=0.5$.

For comparison, we also included in Fig.~\ref{figure6} the ET
for the case when 
the optimum presents an abrupt jump from $\varphi=0.5$ at $t<t_i$ to
$\varphi=0.5+A$ at $t>t_i$, a situation that could model a {\em catastrophic 
change} in the environment.
The interpretation of the corresponding 
curve ``$III$" is obvious. It shows, for example, that a population
that could survive a rather rapid but smooth variation of the optimum cannot 
survive a catastrophe.

This detailed discussion clearly illustrates the delicate interplay 
between the time-scale and amplitude
of the perturbation, on one hand, and the intrinsic time-scales and
nonlinearities of the unperturbed dynamics, on the other hand. 
This interplay is determinant even in this simplified MF-type of approach; 
we are thus expecting it to play a prominent role
for the case of a more microscopic, individual-based modeling,
for which the fluctuations (in particular, the mutations) 
are enriching the dynamics even further.

\section{IBM-type simulations}
\label{ibm2}

Let us turn now to the results obtained
through numerical simulations 
in the IBM-type of approach described in Sec.~\ref{ibm1}. 
One is apriori entitled 
to expect a richer dynamics in this
microscopic approach, since: (a) on one hand, the mutation 
is definitely playing a role, and  ${\cal M}$
represents the second control parameter 
besides the selection ${\cal S}$; (b) on the other hand, the fluctuations
and the finite size of the system can influence
in a nontrivial way the global dynamics of this highly-nonlinear system.

The initial conditions for the system, as already mentioned in Sec.~\ref{ibm1},
are represented by the positions of the individuals on the nodes of the
lattice, as well as by their individual traits. If not explicitly stated
otherwise, we shall consider hereafter that initially the $N(0)$
individuals are randomly, uniformly distributed on the lattice, 
with a given initial concentration $c(0)=N(0)/L^2$; 
also, their traits are randomly, uniformly
distributed in the interval $[0,\,1]$. The case when the $N(0)$
individuals have the same trait was also studied.
The initial concentration is fixed throughout   
to the rather larg
value $c(0)=0.7$. The role of $c(0)$ on the survival of the population, 
and the existence of a critical initial concentration 
(the so-called {\em minimum viable population concentration}) 
below which the 
population gets extinct for given environmental conditions are 
well-known issues, see e.g.~\cite{shaffer, gilpin, katja}, 
and we shall not address them further here. 
In the foregoing we shall keep  $N_{{\rm off}}=6$.

The control parameters that determine the dynamics of the system are 
the SP --  ${\cal S}$ and the MA --  ${\cal M}$. 
In the case of a variable 
environment, as described by Eq.~(\ref{varopt}), one has to consider two 
further control parameters, namely the amplitude $A$ and period 
$T_p$ of the perturbation  of the optimum. The onset 
time  of the optimum's variation is fixed to
$t_{{\rm i}}=1000$ MCS, 
and it is sufficiently large 
such that the system relaxes to a stationary state
in the interval between $t=0$ and $t=t_i$. 

Even when the above parameters, as well as the initial concentration
and mean trait are fixed, there are still various
possible realizations of the system's evolution,   
that are induced by the different sources of
stochasticity which are present in the system,
namely, (a) the randomness in the initial conditions 
(random positions of the individuals on the lattice, and/or to
their randomly-distributed individual traits);
(b) the stochastic aspect in each individual's death;
(c) the random, diffusion-like jumps of the individuals on the lattice;
(d) the mating and the stochasticity elements in the progeny-birth, including the
mutation in the trait of the resulting offsprings.

Some measurable quantities [(i) to (iv) below]
that can characterize in an efficient way the 
behavior of the populations can be cast into two categories: 
those referring to a single realization of the population's evolution, 
and those related to an ensemble
of such realizations.

For a {\em single realization}, these
quantities are, respectively:\\
(i) The ``global" concentration 
$c(t)={N(t)}/{L^2}$, where $N(t)$ is the number of individuals  at
time $t$.\\ 
(ii) The distribution of the individual traits $P(z,t)$ and, 
in particular, the time-dependent value of the mean trait, 
\begin{equation}
\langle z(t)\rangle = \int_0^1 dz\, z\,P(z,t) =
\frac{1}{N(t)}\sum_{i=1}^{N(t)}z_i\,. 
\label{meantrait}
\end{equation}
Another related quantity, currently used by biologists \cite{holt},
is the {\em mean maladaptation},
\begin{equation}
\langle \mu (t) \rangle = \int_0^1 dz\, |z-\varphi(t)|\,\,P(z,t)=
\frac{1}{N(t)}\sum_{i=1}^{N(t)}|z_i-\varphi(t)|\,,
\label{meanmala}
\end{equation}
which represents a measure of the deviation of the
population from the instantaneous optimum imposed by the 
habitat.\\
(iii) The distribution of the individual extinction probabilities
$\Pi(p,t)$, that is directly related to $P(z,t)$. 
Indeed, note that $\Pi(p,t)|dp|=P(z,t) |dz|$, from which one finds,
using Eqs.~(\ref{surv}) and (\ref{fitness}) that:
\begin{equation}
\Pi(p,t)=\frac{{\cal S}}{(1-p) \,\ln^2 (1-p)}P(z,t)\,.
\label{ppi}
\end{equation}

For an {\em ensemble of realizations}, the main element 
of interest is represented by:\\
(iv) The mean extinction time (MET) $\overline{t_{{\rm ex}}}$
for a fixed set of parameters. 
Indeed, due to the intrinsic randomness
of a population's evolution, the extinction time is a
stochastic variable, that differs from one realization of the dynamics
to another. As such, one can consider the associated probability
distribution funtion of the
extinction time, ${\cal T}(t_{{\rm ex}})$ and, 
in particular, its first moment,
\begin{equation} 
\overline{t_{{\rm ex}}}=\int_0^{\infty} dt_{{\rm ex}}\, t_{{\rm ex}}\,
{\cal T}(t_{{\rm ex}})\,. 
\end{equation} 

\subsection{Constant environment}
\label{ibm2const}

Let us consider first a situation that is analogous to 
the one described in Sec.~\ref{mf2const}, namely  a 
constant environment with $\varphi=0.5$, 
in which evolves a finite population with random 
initial conditions, which means random positions of the 
individuals on the nodes of the lattice, as well 
as randomly distributed individual traits with
$\langle z (t=0)\rangle =0.5$.

The main question is whether the phase transition
``extinct-alive" encountered in the MF approach
is destroyed by the fluctuations, or
is it still appearing in this IBM description.  
The situation is of course more complicated, due 
to the effects of the mutation, i.e., due to the existence of the 
second control parameter represented by the amplitude ${\cal M}$
of the mutation (see Sec.~\ref{ibm1}). A related question is, naturally,
how is the mutation affecting the appearance and location of a phase transition.

\subsubsection{Evolution of the concentration $c(t)$}

An indication on a possible phase transition is obviously offered by
the asymptotic behavior of the concentration $c(t)$ 
as a function of the selection parameter ${\cal S}$,
for given initial conditions and a fixed value of ${\cal M}$. 
Indeed, (a) for  ``large-enough" selection pressures ,
$c(t)$ is decreasing, with random fluctuations, till 
the absorbing $c=0$ state; (b) for  ``small-enough" values of ${\cal S}$,
in the long-time limit the concentration $c(t)$ reaches
a regime of stationary fluctuations around a non-zero constant
mean value $c^S$. These results 
clearly point to the existence of a phase transition ``extinct-alive".
Of course, one has to be aware that a ``true" phase transition 
can only appear in the thermodynamic limit 
of an infinite population on an infinite-size lattice, 
see also~\cite{finitesize}. 
In particular, due to the finiteness
of the system, the transition is not abrupt (i.e., there is not
a clear-cut critical value ${\cal S}_c$),
and the frontier between the basins of attraction 
of the absorbing state $c=0$
and of the non-trivial state $c^S$ is also spread out.

\subsubsection{Behavior of the MET  $\;\overline{t_{{\rm ex}}}$}
\begin{figure}
\psfrag{sel}{{\large${\cal S}$}}
\psfrag{TTT}{\large{$\overline{t_{{\rm ex}}}$}}
\psfrag{M = 0}{{\large$\hspace{-0.1cm}{\cal M} =0$}}
\psfrag{M = 0.05}{{\large$\hspace{-0.1cm}{\cal M}=0.05$}}
\psfrag{M = 0.1}{{\large$\hspace{-0.1cm}{\cal M}=0.1$}}
\psfrag{M = 0.5}{{\large$\hspace{-0.1cm}{\cal M}=0.5$}}
\begin{center}\quad \vspace{0.5cm}\\
\includegraphics[width=0.9\columnwidth]{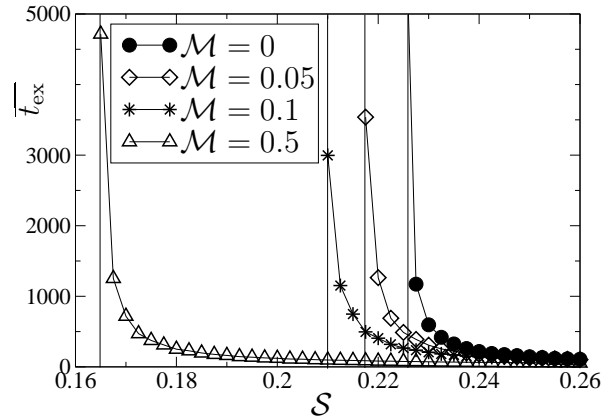}
\caption{\label{IBM7} 
IBM approach: The mean extinction time $\overline{t_{{\rm ex}}}$ 
in a constant environment
as a function of the selection parameter ${\cal S}$, 
for different values of the mutation ${\cal M}$.
The vertical lines indicate the estimated locations of the 
critical selections
for each value of ${\cal M}$.
The other parameters are
$c(0)=0.7$, $L=100$, $\varphi=0.5$, and the average was 
taken over $100$ realizations.}
\end{center}
\end{figure}

Another good indication about the existence 
of the phase transition ``extinct-alive" is  offered by the 
behavior of the MET $\overline{t_{{\rm ex}}}$
as a function of  ${\cal S}$.  
The main results are resumed in Fig.~\ref{IBM7}, that 
indicates a sudden, very abrupt increase in 
$\overline{t_{{\rm ex}}}$
with decreasing ${\cal S}$, which is the signature of 
the phase transition.
Moreover, Fig.~\ref{IBM7}
is illustrating the way the mutation 
${\cal M}$ modifies the location of the transition.
  
\subsubsection{Role of the mutation}

It appears from Fig.~\ref{IBM7} that  increasing mutation amplitude  
${\cal M}$ is unfavorable to 
the survival of a population in a constant environment
and the transition ``extinct-alive" is shifted to lower values of ${\cal S}$. 
The argument  
is that a larger ${\cal M}$ implies a
larger chance for an offspring to be ill-adapted to the constant environment,
even though the parents might be well-adapted. 
\begin{figure}
\psfrag{sel}{{\large${\cal S}$}}
\psfrag{M = 0, t = 100}{{\large$\hspace{-0.1cm}{\cal M} =0,\,t=100$}}
\psfrag{M = 0, t = 1000}{{\large$\hspace{-0.1cm}{\cal M} =0,\,t=1000$}}
\psfrag{M = 0.1}{{\large$\hspace{-0.1cm}{\cal M} =0.1$}}
\begin{center}\quad \vspace{0.5cm}\\
\includegraphics[width=0.9\columnwidth]{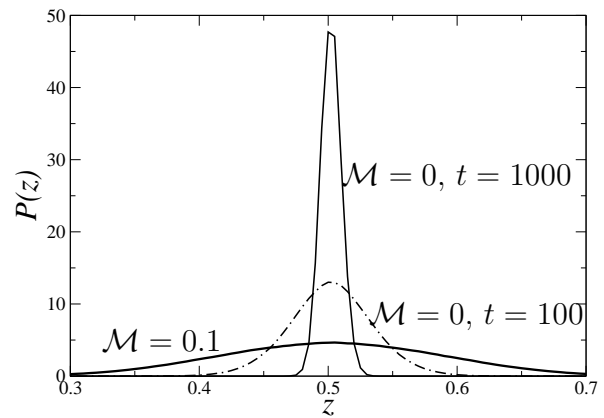}
\caption{\label{IBM10} 
IBM approach: Snapshots of the probability 
distribution function of the trait $P(z,t)$ at
two times $t=100$ and $t=1000$. The thin continuous and dashed-dotted 
lines refer to the 
population without mutation, ${\cal M}=0$. 
The thick line corresponds
to the population with ${\cal M}=0.1$, for which
$P(z,t)$ reaches after some $10$ MCS
a stationary profile.  The values of the other 
parameters are ${\cal S}=0.13$, $c(0)=0.7$, $L=4000$ (for a good statistics),
and $\varphi=0.5$.}
\end{center}
\end{figure}

To get a more quantitative picture of the various factors that 
are involved in this result let us consider a given population 
and focus on the behavior of the probability distribution function
of the trait $P(z,t)$~\footnote{In view of the one-to-one 
relationship~(\ref{ppi}) between $P(z,t)$ and $\Pi(p,t)$, 
one can, alternatively, focus on the probability distribution function of the 
individual extinction probability, 
$\Pi(p,t)$.}. Let us fix the value of the selection -- for example, ${\cal S}=0.165$,  
such that, according to Fig.~\ref{IBM7}, a population 
without mutation (${\cal M}=0$)
will survive (i.e., its asymptotic regime will
correspond to a constant mean concentration $c^S \neq 0$).
Let us compare it with the case of a population 
with non-zero mutation ${\cal M}=0.1$, i.e., a population that, 
again according to Fig.~\ref{IBM7},
will not survive the SP. 
Then, according to Fig.~\ref{IBM10}, for the population without
mutation, ${\cal M}=0$, the distribution of the trait $P(z,t)$ 
is narrowing towards
a $\delta$-peak around $z=\varphi=0.5$ by a progressive elimination,
due to the SP,
of the ill-adapted individuals. Moreover, the offsprings of
well-fitted parents will be also well-fitted, since their trait is
a pure inheritance equal to the mean trait of the parents. 
On the contrary, for the case of the population with ${\cal M}=0.1$
one notices that, after a rather rapid transient
(a few 10s MCS), the distribution of the traits 
remains practically constant. So, although 
the ill-adapted individuals are eliminated in priority by the 
SP (the distribution of the traits tends to be 
narrowed by the selection),
at each mating the traits of the offsprings are randomized again by the 
mutation (the distribution of the traits tends to be 
broadened by the mutation). 
The resulting stationary distribution, that has a finite width 
of the order of ${\cal M}$, is a balance between these two tendencies.

Therefore the population with ${\cal M}=0.1$ is at each moment more 
vulnerable 
than the population with ${\cal M}=0$, and may thus 
die even if the population without
mutation is surviving.

One notices that the evolution of the population, even in the case ${\cal M}=0$,
is much more complex than that described by the MF approach. In
particular, the dynamics of the concentration and of the distribution of the
individual traits  are strongly 
inter-related. As long as ${\cal M}=0$, they are taking place on a comparable time
scale, but it is difficult to assimilate this time scale with the one
of the simple relaxational processes that appear in the 
MF description. However, when
${\cal M}\neq 0$, the evolution of the concentration 
and that of $P(z,t)$   
take place, in general,  on different time
scales. For example, for the situation of a population with ${\cal M}=0.1$
considered in Fig.~\ref{IBM10} , the relaxation of $P(z,t)$ 
to its asymptotic profile takes place during a 
short transient time of the order of $10^2$ MCS, while the 
relaxation of the concentration $c(t)$ to the absorbing state $c=0$
is very slow, with a characteristic time of the order 
$10^5$ MCS, that can be estimated from the 
asymptotic part of the plot $c(t)$. 

As can be seen, the mutation
is introducing essentially new features in the dynamics, even in the
simplest context of a constant environment. In
particular, the characteristics of the population (like concentration, trait
and individual extinction probabilities) can evolve on different time scales,
that are sensitive to both control parameters of selection ${\cal S}$ and 
mutation ${\cal M}$. This fundamental role of the mutation is not 
at all accounted for in the  MF description.

After discussing these essential  {\em qualitative differences} that appear
between the MF and the IBM results, let us turn to the problem of
the calibration of the parameters corresponding to the two levels of modeling.
One has to retain that our MF description is not the result of a rigorous 
coarse-graining procedure applied to the IBM-model.
As such, besides the problem of the correspondence of the unit-time already mentioned
in Sec.~\ref{model}, the control parameter of selection ${\cal S}$ that appears in
both approaches may present some calibration problems, too. 
Moreover  the difficulty in establishing 
a quantitative correspondence
between the two aproaches is actually far from being a simple 
rescaling problem.It is related to the role of the 
spatial correlations between the individuals 
(like, for example, possible clustering effects),
role that is completely neglected in the MF approach.
Let us consider just two aspects  
in order to illustrate this point: (a) It seems that these correlations 
tend to increase the value of the asymptotic
concentration $c^S$, however they are contributing  
to reduce the survival chances
at increasing SP as compared to the MF. 
(b) The comparison between the role that 
the maximum number of offsprings $N_{{\rm off}}$ is playing
in the MF description and in the IBM one indicates that
in both cases the critical value of the selection ${\cal S}_c$ is
increasing with $N_{{\rm off}}$. However, the concentration $c^S$ of the 
stationary non-absorbing state is decreasing with $N_{{\rm off}}$ in the
MF description, while it is increasing with 
$N_{{\rm off}}$ in the IBM approach. 

\subsection{Time-periodic changes of the environment}

Let us now turn to the case of a time-dependent habitat, whose optimum 
varies in time according to Eq.~(\ref{varopt}).

\subsubsection{Phase diagram: role of selection ${\cal S}$ and mutation ${\cal M}$}

\begin{figure}[h!]
\psfrag{sel}{{\large${\cal S}$}}
\psfrag{lll}{$\varphi=0.5$}
\psfrag{mmm}{$\varphi=0.8$}
\psfrag{M}{{\large${\cal M}$}}
\begin{center}\quad \vspace{0.75cm}\\
\includegraphics[width=1.\columnwidth]{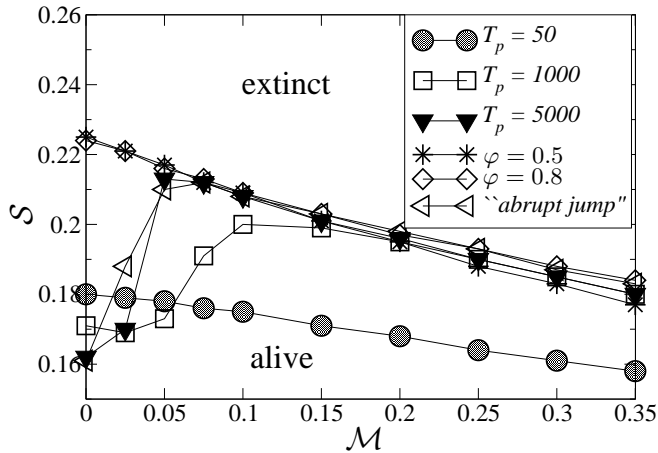}
\caption{\label{IBM13} 
IBM approach: The phase diagram ``extinct-alive" in the  ${\cal S}$ and ${\cal M}$  plane, for various 
periods $T_p$ of the optimum oscillation. For comparison we also presented the
results for a constant environment with $\varphi=0.5$,
respectively $\varphi=0.8$, as well as for an abrupt jump
of the optimum (at $t_i=1000$) between these two values.  
The average was done over 100 runs and the vertical 
size of the symbols is of the order of the estimated error
bars. The continuous lines are just a guide for the eye. 
The values of the other parameters
are $A=0.3$, $c(0)=0.7$, and $L=100$.}
\end{center}
\end{figure}

Consider the location of the phase transition point ``extinct-alive", i.e., 
the critical value of the SP for a  population with given initial conditions.
This critical point depends not only on the amplitude $A$ and period $T_p$
of the optimum variation, but also on   ${\cal M}$ --
an aspect which of course cannot be captured by the MF description. 
One is thus led to consider the phase diagram in the plane of the 
characteristic parameters  ${\cal S}$ and ${\cal M}$, 
see Fig.~\ref{IBM13}.   

Several characteristics are emerging from the analysis 
of these critical curves,
namely:\\

\noindent(a) The phase diagram is very little affected by the 
actual value of the optimum $\varphi$ of a {\em constant} environment. 
Indeed, as illustrated in Fig.~\ref{IBM13},  the curves for
$\varphi=0.5$ and $\varphi=0.8$ present little differences. 

This result holds provided that the population has initially
randomly-distributed individual traits. In this case,
whatever the constant value of the optimum $\varphi$, a fraction 
of the initial population is (more or less) ill-adapted and 
will be progressively 
eliminated. Also, the descendancy will become progressively
more and more fitted to the environment.
As such, in the long run the distribution of the trait of the 
population will become peaked around $\varphi$, with a variance 
of the order of ${\cal M}$, and the corresponding distribution of the 
individual extinction probabilities will be essentially the same 
for all constant values of $\varphi$. The small differences that appear,
especially for small values of ${\cal M}$, in
the critical curves for $\varphi=0.5$ and $\varphi=0.5+A$
are due to the way we prescribed the trait of a 
progeny, according to Eq.~(\ref{zoff}).
When starting from randomly-distributed initial 
individual traits, the heredity part 
tends to ``push" the descendancy towards 
a mean trait equal to $0.5$; thus the relaxation towards the 
asymptotic profile of the distribution of the individual 
extinction probabilities will be slower 
for $\varphi=0.5+A$ than for $\varphi=0.5$. This makes 
the population with $\varphi=0.5+A$ a little bit more sensitive to the SP.

Also, as already mentioned in the previous section, for a constant environment
any modification of the {\em status quo} of the population due to the mutation 
is ``bad", and the
critical value of ${\cal S}$ decreases monotonously
with ${\cal M}$.\\

\noindent (b) Consider now a population with initially equal 
individual traits $z_i=0.5$ and put it in a constant environment
with $\varphi=0.5+A$. 
For small values of the mutation ${\cal M}$ the randomness in the 
descendancy traits cannot overcome the misfit inherited 
from the parents. It is thus clear that the population will get extinct
for smaller values of ${\cal S}$  than in the cases
discussed above. This point
is also reflected in the behavior of the MET.
The result for ${\cal M}=0$  is in agreement with the
MF case, see Fig.~\ref{figure6}.

However, for  large values of ${\cal M}$ the inherited traits of the 
offspring are greatly changed by mutations and the progeny adapts
to the constant environment with $\varphi=0.5+A$. The critical curve
is approaching the ones for the constant optimum discussed at point
(a) above. 

For intermediate values of ${\cal M}$, inheritance and mutation 
are both important in
the offspring traits, and the critical curve still lies below the ones
corresponding to point (a). This is in agreement with the behavior of 
the mean extinction 
time, as illustrated in Fig.~\ref{IBM18} for ${\cal M}=0.1$.

As such, the population is sensitive to the initial distribution of the
individual traits only for small values of ${\cal M}$, and the memory
of the initial state is lost when mutation is strong.\\

\noindent (c) Any variation in the optimum is harmful for the system,
as indicated by the fact that all the critical curves corresponding
to a variable optimum lie below the one for the constant optimum
$\varphi=0.5$.
This is again in qualitative agreement with the MF results.\\

\noindent (d1) Consider the case of a 
{\em population without mutation, ${\cal M}=0$}.
The less favorable situation for such a population 
seems to be that of an abrupt jump in the optimum
from $\varphi=0.5$ to $\varphi=0.5+A$, 
as compared to all the cases of smooth
oscillation of the optimum around $\varphi=0.5$. This result is 
in {\em qualitative} agreement with the one in Ref.~\cite{andre02}, 
and can be understood as follows: at $t=t_i$, when the perturbation 
of the optimum sets in, for a population with ${\cal M}=0$ the
overhelming majority of the individuals (and thus their children)
have an individual trait $z_i=\varphi=0.5$. 
An abrupt change in the optimum $\varphi$ 
will lead to an abrupt increase in the extinction probability 
of the individuals (per MCS) till its maximum possible value
$p=1-\exp[-{\cal S}/(1-A)]$, corresponding to the given amplitude $A$.
This leads  to a decrease 
in the critical  value of the selection ${\cal S}$. 
One can therefore conclude that after a sharp change 
in the living conditions, such as a large-scale catastrophe, 
even a moderate SP could be lethal
for a population lacking the variety induced by mutation.

Moreover, a long period of the optimum variation is less favorable 
to the population survival than a shorter one. Indeed, one can argue 
that in the case of a large period the individuals are spending a 
longer time (in MCS) in an unfavorable environment, and thus have 
more chances to die. Recall that the absence of mutation leads to the 
fact that the descendancy is as ill-adapted to the environment as the parents are. 

\begin{figure}[h!]
\psfrag{X}{{\large${\cal S}$}}
\psfrag{M = 0}{{\large$\hspace{-0.15cm}{\cal M} =0$}}
\psfrag{M = 0.05}{{\large$\hspace{-0.15cm}{\cal M}=0.05$}}
\psfrag{M = 0.1}{{\large$\hspace{-0.15cm}{\cal M}=0.1$}}
\psfrag{M = 0.5}{{\large$\hspace{-0.15cm}{\cal M}=0.5$}}
\psfrag{TT}{\large{$\hspace{-0.2cm}\overline{ t_{ex} }$}}
\begin{center}\quad \vspace{0.75cm}\\
\includegraphics[width=0.9\columnwidth]{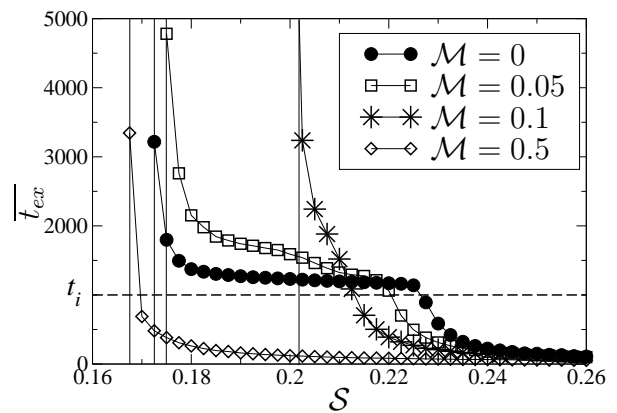}
\caption{\label{IBM16} 
IBM approach: The mean extintion time $\overline{t_{ex}}$ 
as a function of
t ${\cal S}$ for different values of ${\cal M}$. 
The thin vertical lines indicate the estimated location of the transition point.
The average was
done over 100 populations with $c(0)=0.7$,
$L=100$, $A=0.3$, $T_p=1000$ and $t_{{\rm i}}=1000$.
The continuous lines are just a guide for the eye.}
\end{center}
\end{figure}

These elements are also illustrated in Fig.~\ref{IBM16} for the dependence of
the mean-extinction time on ${\cal S}$.\\

\noindent (d2) Consider now a population with an {\em intermediate
value of the MA}, say ${\cal M} \approx 0.1.$ 
One notices that, contrary to the situation (d1) of a population with ${\cal M}=0$,
an increase in the period $T_p$ of the optimum change 
is beneficial for the population and the critical  
SP is deplaced towards larger values.

Indeed, for the case of an intermediate ${\cal M}$, the heredity 
and the mutation are equally important in establishing the
trait of the population. As such, if the optimum is varying slowly, 
the population has the time to adapt to the instantaneous value
of the optimum, and ``good traits" are transmitted, through heredity, 
to the descendancy; moreover, the randomness element due to the mutation
is somehow compensating for the small variation of the optimum from one
generation to the other. 
On the contrary, if the optimum is varying rapidly from one generation 
to the next one, even if the parents were well-adapted at some
instantaneous value of $\varphi$, they will become soon ill-adapted
and transmit this misfit to their progeny through heredity.
The mutation element in the offspring traits cannot compensate for the 
too large inherited misfit between the individual trait and 
the instantaneous optimum.
 
Therefore, a population with an intermediate mutation amplitude is more
vulnerable to rapid variations in the optimum. This result is also 
illustrated in Fig.~\ref{IBM18} through the behavior of the mean extinction
time of a population, as a function of ${\cal S}$, for different values of the
optimum period $T_p$.\\

\noindent (d3) {\em Very large MA} is always harmful
for the population's survival, 
whatever the period of change $T_p$ of the optimum. Indeed,
a large MA changes so much the traits that the 
inheritance of good traits tends to be lost. 
Each generation is therefore composed of too many individuals with
unfitted traits (since the ``cleaning action" that the selection 
operated on the previous generations does not lead to a better-fitted 
offspring). The critical value of ${\cal S}$ is decreasing 
monotonously with ${\cal M}$ in the whole region of high MA. \\

\noindent (e) The small-$T_p$ curves are, in general, monotonously 
decreasing with increasing ${\cal M}$, suggesting that 
larger mutations, combined with rapid changes of the environment, 
are harmful for the population, i.e., reduce its 
resistance to the SP.\\

\noindent (f) Finally, the most spectacular element of this phase diagram
is the {existence of an optimal value of
the mutation} for intermediate values of $T_p$ (at fixed $A$), 
i.e., a maximum in the
critical curve ${\cal S}$--${\cal M}$. For this {\em optimal mutation}, 
the critical value of the selection is maximum, 
indicating that the population resists to higher
SPs than for other values of ${\cal M}$.
The value of this optimal mutation depends on the period $T_p$ 
of the optimum variation.

A crude explanation of this effect would follow these lines:
Consider first a population with a small MA,
that brings diversity to the phenotypic pool of the population. 
Through the selection, unfitted individuals tend to be eliminated, 
and the fitted ones survive. If the optimum varies slowly (large $T_p$), 
then the ``good" traits, that will be passed to the next
generations, plus a small mutation, remain good for some time. 
A small amount of mutation is therefore
beneficial. If, however, $T_p$ is too short, then quite soon the previously
good traits (that a too small mutation cannot correct, in order to accomodate
them to the evolving value of the optimum !)
turn out to be bad. A larger effect of the mutation is needed
to ensure an optimal adaptation of the descendancy to the new
values of the optimum; however, see (d3), the MA should not be ``too large" 
either. Such effects of a beneficial range of mutation 
have already been observed in biological systems, see e.g.~\cite{zawierta}.

This discussion also indicates that an optimal mutation can only 
be found for intermediate values of $T_p$; too large or too small
rates of changes of the optimum do not allow for the above-described 
effects, as illustrated in Fig.~\ref{IBM13}.
The MET can be thus 
increased by the optimal amplitude of the mutation, for a given 
period $T_p$ of the optimum variation. This result is illustrated in
Fig.~\ref{IBM16} for the case of $T_p=1000$, for which the phase diagram
of Fig.~\ref{IBM13} indicates an optimal value of the mutation
${\cal M} \approx 0.1$. Indeed, for values of ${\cal M}$ 
above and below this opimal value,
the MET (for each value of ${\cal S}$) is clearly lower
than the one corresponding to the optimal mutation. 

The signature of the phase transition ``extinct-alive" 
can be also found in the behavior of the profile
of the probability distribution function of the extinction time
${\cal T} (t_{{\rm ex}})$ for an ensemble of populations with fixed 
initial conditions $c(0)$, $\langle z\rangle$ and given parameters 
${\cal M}$, $A$, and $T_p$.
Indeed, below the critical value of ${\cal S}$ the distribution of 
the extinction time is narrow-peaked around 
$\overline{t_{{\rm ex}}}$, while above the transition its
profile changes, by developing a fat tail towards 
large values of $t_{{\rm ex}}$. Therefore, above the critical point 
$\overline{t_{{\rm ex}}}$
is no longer corresponding to the most representative value of 
$t_{{\rm ex}}$.

\subsubsection{Role of the period $T_p$ and amplitude $A$ 
of optimum oscillations}

The influence of the period $T_p$ (for a fixed value of $A$)
on the critical value of  ${\cal S}$ 
depends strongly on the value of  ${\cal M}$. 
As shown in  the previous 
paragraph, for  very small or zero values of ${\cal M}$, 
a decrease in $T_p$ (i.e., slow perturbations of the optimum) 
leads to an increase in
the critical value of ${\cal S}$. 

\begin{figure}[h!]
\psfrag{XXX}{{\large${\cal S}$}}
\psfrag{M}{{\large${\cal M}$}}
\psfrag{TTT}{\large{$\hspace{-0.2cm}\overline{ t_{ex} }$}}
\psfrag{L05}{$\varphi\,=\,0.5$}
\psfrag{L08}{$\varphi\,=\,0.8$}
\begin{center}\quad \vspace{0.75cm}\\
\includegraphics[width=0.9\columnwidth]{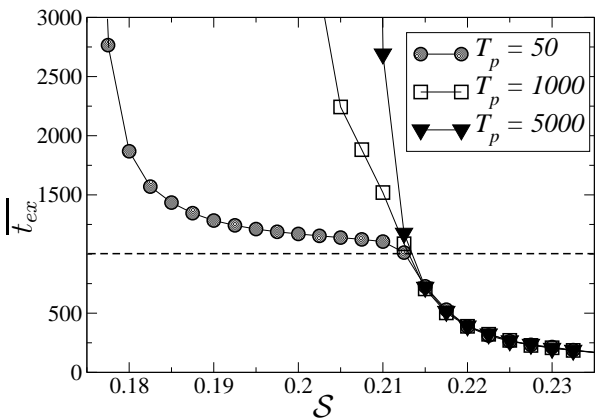}
\caption{\label{IBM18} 
IBM approach: The mean extintion time $\overline{ t_{ex} }$ as a function of
the selection pressure ${\cal S}$ for ${\cal M}=0.1$ (intermediate MA)
and different values of $T_p$.
The results are averages  
done over 100 populations with $c(0)=0.7$,
$L=100$, $A=0.3$, and $t_{{\rm i}}=1000$.}
\end{center}
\end{figure}

However, for intermediate values of ${\cal M}$, 
the role of $T_p$ gets reversed, namely
increasing $T_p$ (with fixed $A$) leads 
to an increase in the critical ${\cal S}$.  
This point was 
 discussed in the previous paragraph,
and it is shown in Fig.~\ref{IBM18} of
$\overline{ t_{ex} }$ versus ${\cal S}$ for ${\cal M}=0.1$.

Finally, for the case 
of a population with a large MA,
the MET is less sensitive to the period $T_p$ of
the optimum oscillation, as well as to the initial distribution of the
individual traits than in the case of intermediate or small values of
${\cal M}$.

\begin{figure}[h!]
\psfrag{X}{{\large${\cal S}$}}
\begin{center}\quad \vspace{0.75cm}\\
\includegraphics[width=0.9\columnwidth]{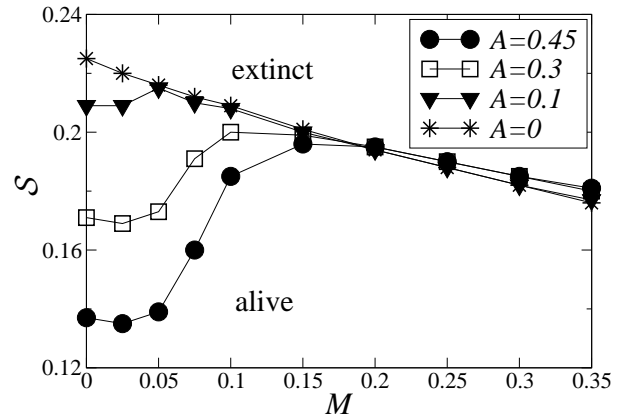}
\caption{\label{IBM19} 
IBM approach: Role of the amplitude $A$ of the optimum oscillation
on the phase diagram in the (${\cal S}$, ${\cal M}$) plane.
The average  was
done over 100 populations with $c(0)=0.7$,
$L=100$,  $t_{{\rm i}}=1000$, and the period of 
optimum oscillation is
$T_p=1000$. 
The continuous lines are just a guide for the eye.}
\end{center}
\end{figure}

The role of the oscillation amplitude $A$ 
is easy to resume: increasing $A$ 
has a destabilizing effect on the system, i.e., it deplaces 
the critical curve towards smaller values of the SP,
see Fig.~\ref{IBM19},
and it reduces the extinction time of the population
(all the other parameters, as well as the initial conditions being kept the
same). This effect, however, is more or less pronounced depending on the values
of the other parameters $T_p$ and ${\cal M}$. 
In particular, for large 
MAs ${\cal M}$ the transition 
point depends only weakly on the value of $A$. 

\subsubsection{Influence of the system size}

The system size is playing a role on the 
location of the transition point, as well as on its ``sharpness", 
see Ref.~\cite{finitesize}.
As expected on general backgrounds, 
the width of the transition zone is increasing with decreasing system size.
Moreover, small-size populations are, on average, 
more vulnerable than larger-size
ones, i.e., it is more probable that small-size systems 
get extinct for
smaller values of the SP 
than the analogous large ones.
This effect is essentially
related to the larger amplitude of the relative fluctuations in the
number of individuals in smaller-size systems as compared to 
larger-size ones; indeed,  for a given mean concentration, this amplitude 
goes roughly as $1/L$.
This effect is well-known in biology as the
{\em demographic stochasticity effect} in population extinction, see 
Refs.~\cite{mdap2,shaffer} for further comments.

\subsubsection{Evolution of the concentration $c(t)$}

\begin{figure}[h!]
\psfrag{X}{$\langle z  (t) \rangle$}
\psfrag{Y}{$\varphi (t)$}
\psfrag{mala}{$\langle \mu (t) \rangle$}
\begin{center}\quad \vspace{0.75cm}\\
\includegraphics[width=0.9\columnwidth]{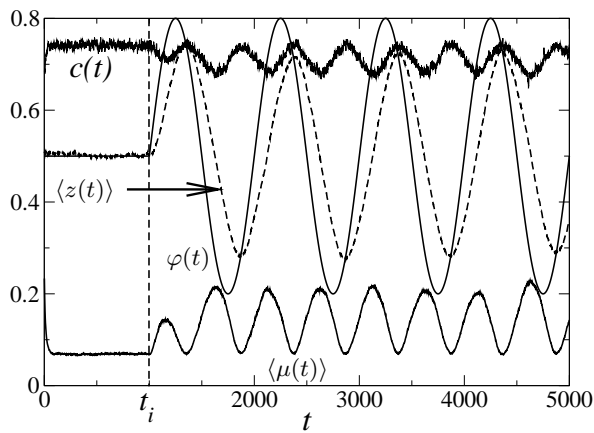}
\caption{\label{IBM20} 
IBM approach: The oscillations of the concentration $c(t)$, mean trait $\langle
z (t) \rangle$ (dashed line), and mean maladaptation $\langle \mu (t) \rangle$
for a population that survives in an oscillating habitat.
The values of the other parameters
are ${\cal M}=0.1$, ${\cal S} = 0.13$, $A=0.3$, $T_p=1000$, 
$t_{\rm{i}}=1000$, $L=100$, and $c(0)=0.7$. For comparison we also represented
the oscillating optimum $\varphi(t)$.}
\end{center}
\end{figure}

For ``large" (i.e., above the transition point)
values of the SP the concentration
$c(t)$ tends to zero, but below the transition point
the concentration reaches, after a transient regime, an
oscillatory behavior of period $T_p/2$, as illustrated in 
Fig.~\ref{IBM20}.
Of course, due to the inherent stochastic nature of the dynamics, these
oscillations are noisy.
As in the MF case, the amplitude of these oscillations
(for given ${\cal S}$ and ${\cal M}$) depends on both $A$ and $T_p$.
The oscillations 
of the concentration are not in phase with the absolute value of the 
optimum, but  exhibit a mean lag that depends on the control parameters
$A$, $T_p$, ${\cal S}$, and ${\cal M}$ of the system. In 
Fig.~\ref{IBM20} we also represented the temporal evolution of the
mean population trait $\langle z (t)\rangle$, see
Eq.~(\ref{meantrait}). Its oscillations, of period $T_p$,
appear only for ${\cal M}\neq 0$, while
for ${\cal M}=0$ the mean trait settles down to a constant value
that is equal the mean value of the optimum ($\langle z \rangle = 0.5$).
Finally,  Fig.~\ref{IBM20} presents the oscillations 
(of period $T_p/2$)
of the mean maladaptation  $\langle \mu (t)\rangle$ (Eq.~(\ref{meanmala})), 
and one notices that an instantaneous minimum in the concentration 
corresponds to a maximum of the maladaptation. Again, the characteristics of these
curves (amplitude, mean over one period, lag as compared to the optimum
variation, etc.) depend on ${\cal S}$ and ${\cal M}$.


\section{Conclusions and perspectives}

In this paper we have considered a model 
of single-species population dynamics 
on a $2D$ lattice,
and we analyzed the role of the selection pressure 
and mutation on 
its behavior in a periodically-changing environment.
Two levels of description were considered in detail  --  a
mean-field one, and  an individual-based approach with built-in 
stochasticity. 

The MF approach allowed us to highlight, 
in a rather intuitive manner, the delicate
interplay between the different time-scales of the processes 
involved in the
dynamics of this highly-nonlinear system. 
In particular, the influence of the amplitude and periodicity
of the optimum variation on the critical value of the SP
(above which the population gets extinct) was discussed in  detail.
However, this type of MF approach is unsuitable for describing another
essential element of the population dynamics, namely the 
stochasticity introduced by the mutation in the descendancy 
traits~\footnote{Adaptive dynamics (see e.g.~\cite{diekmann} 
for an introduction) studies how a population
of agents having all the same trait reacts to the presence of an individual
(a ``mutant") having a different trait. The associated dynamics is described by
mean-field like partial differential equations. However, in our case, the
generic situation is not of this type, as the traits of the agents are
randomly distributed.}.
 
In order to take into account the effects of the mutation, one has to 
appeal to a more refined level of description like the individual-based one,
in which the stochastic aspects of the dynamics are fully accounted for.
As expected, at the IBM level of description the dynamics of the 
population is richer. In particular, the MA
is influencing in a
highly nontrivial manner the critical properties of the system at the 
phase transition
point between ``alive" and ``extinct". A spectacular effect in the phase
diagram of the system in the plane of the parameters $\cal{S}$--$\cal{M}$ 
(for fixed values of the period and amplitude of the 
environmental changes) is the 
existence of an  optimal MA, i.e., a value of ${\cal M}$ for which the
critical SP is maximal.
The existence of this optimal MA, that is specific to each 
intermediate value of $T_p$, is strongly reminiscent of the 
{\em stochastic resonance} 
phenomenon encountered in stochastic systems submitted to a periodic
deterministic perturbation, see e.g.~\cite{SR}. In a broad sense, this effect
means that the response of a nonlinear dynamical system to a periodic
perturbation of one of its parameters can be enhanced by an optimal amount of
noisy stimulation of the system. In our case, the response of the system is
encoded in the MET, and the noisy ingredient is the random
mutation of amplitude ${\cal M}$. A more explicit, quantitative mapping of 
our system on a nonlinear
stochastic equation for the concentration that exhibits the stochastic
resonance phenomenon is currently under study.
 
We also compared the effects of a smooth variation of the optimum
with those corresponding to an abrupt changing of the environment (as, for
example, a large-scale catastrophic event). We showed that sufficiently
large  mutations can increase dramatically the survival chances 
of the population in case of a catastrophe.

Variations of the model introduced above, for example 
using different expressions
for the individual fitness, or for 
the dependence of the probability of extinction upon
the SP, and fitting better some possible experimental data,
could also be considered. Also, more complicated than diffusive types of 
motion, that may pertain to a 
``strategy" of the individuals (e.g., a tendency to approach or, 
on the contrary, to avoid other individuals) are 
currently under study. We believe however that most of the qualitative
features of our model, resulting from the competition of several basic
processes, are generic.  Extensions to ecosystems composed of several
competing species and submitted to a changing environment is presently
under investigation. 

\acknowledgments{The authors thank an anonymous referee for very 
useful suggestions. 
M.D. and I.B. acknowledge partial support from 
the Swiss National Science Foundation. 
M. D. and J. S. acknowledge the COST10-SER-No.C06.0027
program for support.
A.P. is grateful for the  hospitality of the Theoretical 
Physics group of the University of Geneva.}

\end{document}